\documentclass[aps,pre,twocolumn,superscriptaddress,showkeys]{revtex4-1}
\usepackage{graphicx,color,colortbl}
\usepackage{epstopdf}
\usepackage{bm}
\usepackage{eucal}
\usepackage{mathrsfs}

\usepackage[dvipsnames]{xcolor}
\usepackage{hyperref}
\definecolor{hypcolor}{named}{BlueViolet}
\hypersetup{colorlinks=true, citecolor=hypcolor, urlcolor=hypcolor, linkcolor=hypcolor}

\usepackage{amsmath}
\usepackage{amsthm}
\usepackage{epsfig}
 
\newcommand{\be}{\begin{equation}} 
\newcommand{\ee}{\end{equation}}
\newcommand{\bea}{\begin{eqnarray}} 
\newcommand{\eea}{\end{eqnarray}}
\newcommand{\bes}{\begin{subequations}} 
\newcommand{\ees}{\end{subequations}}

\begin{document}
\title{Population spiking and bursting in next generation neural masses \\ with spike-frequency adaptation}

 \author{Alberto Ferrara}
	\affiliation{Sorbonne Universit\'e, INSERM, CNRS, Institut de la Vision, 75012 Paris, France}
	\author{David Angulo-Garcia}
	\affiliation{Grupo de Modelado Computacional-Din{\'a}mica y Complejidad de Sistemas, Instituto de Matem{\'a}ticas Aplicadas, 
	Universidad de Cartagena, Carrera 6 36-100, Cartagena de Indias 130001, Colombia}
	\author{Alessandro Torcini}
	\affiliation{Laboratoire de Physique Théorique et Modélisation, UMR 8089, CY Cergy Paris Université, CNRS, 95302 Cergy-Pontoise, France}
	\affiliation{CNR, Consiglio Nazionale delle Ricerche, Istituto dei Sistemi Complessi, via Madonna del Piano 10, 50019 Sesto Fiorentino, Italy}		
	\affiliation{INFN, Sezione di Firenze, via Sansone 1, 50019 Sesto Fiorentino, Italy}
    \author{Simona Olmi}
    \email[corresponding author: ]{simona.olmi@fi.isc.cnr.it}
	\affiliation{CNR, Consiglio Nazionale delle Ricerche, Istituto dei Sistemi Complessi, via Madonna del Piano 10, 50019 Sesto Fiorentino, Italy}
	\affiliation{INFN, Sezione di Firenze, via Sansone 1, 50019 Sesto Fiorentino, Italy}
	
	\date{\today}


\begin{abstract}
Spike-frequency adaptation (SFA) is a fundamental neuronal mechanism
taking into account the fatigue due to spike emissions and the consequent reduction of 
the firing activity. We have studied the effect of this adaptation mechanism  on the macroscopic
dynamics of excitatory and inhibitory networks of quadratic integrate-and-fire (QIF) neurons
coupled via exponentially decaying post-synaptic potentials. In particular, we have studied the population
activities by employing an exact mean field reduction, which gives rise to 
next generation neural mass models. This low-dimensional reduction allows for the derivation of
bifurcation diagrams and the identification of the possible macroscopic regimes emerging
both in a single and in two identically coupled neural masses. 
In single popukations SFA favours the emergence of population bursts in excitatory networks, 
while it hinders tonic population spiking for inhibitory ones.
The symmetric coupling of two neural masses, in absence of adaptation, leads to the emergence
of macroscopic solutions with broken symmetry : namely, chimera-like solutions in the
inhibitory case and anti-phase population spikes in the excitatory one. The addition of SFA
leads to new collective dynamical regimes exhibiting cross-frequency coupling (CFC) among the fast
synaptic time scale and the slow adaptation one, ranging from anti-phase slow-fast nested oscillations to
symmetric and asymmetric bursting phenomena. The analysis of these CFC rhythms in the
$\theta$-$\gamma$ range has revealed that a reduction of SFA leads to an increase 
of the $\theta$ frequency joined to a decrease of the $\gamma$ one. This is analogous
to what reported experimentally for the hippocampus and the olfactory cortex 
of rodents under cholinergic modulation, that is known to reduce SFA.
\end{abstract}


\maketitle



\section{Introduction}

Neural mass models are mean field models developed to mimic the dynamics of homogenous
populations of neurons. These models range from purely heuristic ones (as the well-known Wilson-Cowan model \cite{wilson1973}), to more refined versions obtained by considering the eigenfunction expansion of the 
Fokker-Planck equation for the distribution of the membrane potentials  \cite{mattia2002,schaffer2013}.
However, quite recently, a {\it next generation neural mass model} has been derived in an exact manner for
heterogeneous populations of quadratic integrate-and-fire (QIF) neurons \cite{montbrio2015}. This new generation of neural mass models describes the dynamics of networks of spiking neurons in terms of macroscopic variables, like the population firing rate  and the mean membrane potential, and it has already found various applications in many neuroscientific contexts \cite{byrne2017, byrne2020, ceni2020,bi2020,taher2020,segneri2020, gast2020mean, gerster2021,gast2021}.

Neural populations can display collective events, resembling spiking or bursting dynamics, observable at the single neuron level \cite{varona2000,de2006}. In particular, in this context, tonic spiking corresponds to periodic collective oscillations (COs), while a population burst is a relaxation oscillation connecting a  spiking regime to a silent (resting) state  \cite{izhikevich2000}.

Regular collective oscillations  have been reported for spiking neural populations with purely excitatory \cite{van1996,olmi2010} or inhibitory 
interactions \cite{whittington1995}. The emergence of these oscillations have been usually related to the presence of a synaptic time scale
\cite{van1994} or to a delay in the spike transmission \cite{brunel1999}.
Indeed, as shown in \cite{devalle2017firing,ceni2020}, the inclusion of exponentially decaying synapses in inhibitory QIF networks is sufficient for the appearence of COs, corresponding to limit cycles emerging 
via a Hopf bifurcation in the associated neural mass formulation.

A prominent role for the emergence of population bursts is played by spike-frequency adaptation (SFA), a mechanism for which a neuron, subject to a constant stimulation, gradually lowers its firing rate \cite{van2001}. Adaptation in brain circuits is controlled by cholinergic neuromodulation. In particular an increase of the acetylcholine neuromodulator released by
the cholinergic nuclei leads to a clear reduction of SFA in the Cornu Ammonis area 1 (CA1) pyramidal cells of the hippocampus \cite{aiken1995}.
Recently, it has been shown that an excitatory next generation neural mass
equipped with a mechanism of global adaptation (specifically short-term depression or SFA) can give rise to bursting behaviours \cite{gast2020mean}.

Furthermore, cholinergic drugs are responsible for a modification of the neural oscillations frequency, specifically of the 
$\theta$ and $\gamma$ rhythms \cite{traub1992,barkai1994,crook1998}, which are among the most common brain rhythms \cite{buzsaki2006}.
Specifically $\gamma$ oscillations, which have been observed in many areas of the brain \cite{buzsaki2012}, have a range between $\simeq 30$ and $120$ Hz, while $\theta$ oscillations correspond to 4–12 Hz in rodents \cite{colgin2009,belluscio2012}  and to 1–4 Hz in humans \cite{zhang2015}. 
Moreover $\gamma$ oscillations  have been recently cathegorized in three distinct bands for the CA1 of the hippocampus \cite{belluscio2012}: a slow one ( $\simeq 30-50$ Hz), a fast one ( $\simeq 50-90$ Hz), and a so-termed $\varepsilon$ band ( $\simeq 90-150$ Hz). $\gamma$ rhythms with similar low- and high-frequency sub-bands occur in many other brain regions besides the hippocampus \cite{colgin2016, sirota2008} 
and they are usually modulated by slower $\theta$ rhythms in the hippocampus during locomotory actions and rapid eye movement (REM) sleep.
This modulation is an example of a more general mechanism of cross-frequency coupling (CFC) between a low and a high frequency rhythm, which 
is believed to be functionally relevant for the brain actvity \cite{canolty2010}: low frequency rhythms (such as $\theta$) usually involve broad regions of the brain and are
entrained to external inputs and/or cognitive events, while high frequency oscillations (such as $\gamma$) reflect local computation activity. 
Thus CFC can represent an effective mechanism to transfer information across spatial and temporal scales \cite{canolty2010, lisman2013}. 
The most studied CFC mechanism is the phase-amplitude coupling, which corresponds to the modification of the amplitude (or power) of $\gamma$-waves induced by the phase of the $\theta$-oscillations; this phenomenon is often referred as $\theta$-nested $\gamma$ oscillations \cite{butler2016}.
Cholinergic neuromodulation, and therefore SFA, has been shown to control $\theta$-$\gamma$ phase-amplitude coupling in freely moving rats in the medial enthorinal cortex \cite{newman2013}
and in the prefrontal cortex \cite{howe2017}.

In this paper we want to compare the role played by the spike frequency adaptation in shaping the emergent dynamics  
in two simple setups: either purely inhibitory or purely excitatory neural networks. In this regard we consider fully coupled QIF neurons
with SFA, interacting via exponentially decaying post-synaptic potentials.
The spike-frequency adaptation is included in the model via an additional collective afterhyperpolarization (AHP) current, which temporarily hyperpolarizes the cell upon spike emission, with a recovery time of the order of hundreds of milliseconds \cite{traub1992,van2001,gigante2007}.
We will first analyze the dynamics of an isolated population with SFA and then extend the analysis to two simmetrically coupled populations. In the latter case we will focus on the emergence of
collective solutions (either asynchronous or characterized by population spiking and bursting) with particular emphasis of the symmetric or
nonsymmetric nature of the dynamics displayed by each population \cite{ratas2017symmetry}.

The emergence of CFC among oscillations in the $\theta$ and $\gamma$ range have been previously reported for
next generation neural masses: namely, for two asymmetrically coupled inhibitory populations with different synaptic time scales \cite{ceni2020}, as well as for inhibitory and excitatory-inhibitory networks under an external $\theta$-drive \cite{segneri2020}. Here, we show that, in presence of SFA, $\theta$-$\gamma$ CFCs naturally emerge in absence of external forcing and for symmetrically coupled populations. The cross-frequency coupling is due to the presence of a fast time scale associated to the synaptic dynamics and a slow one relative to the adaptation. Quite peculiarly, inhibitory interactions give rise to slow $\gamma$ oscillations (30-60 Hz), while excitatory ones are associated to fast $\gamma$ rhythms (60-130 Hz). Furthermore, we will show that SFA controls the frequency of the slow and fast rhythms as well as the entrainement among $\theta$ 
and $\gamma$ rhythms.

This paper is organised  as follows. Section II is devoted to the introduction of the QIF neuron and of the studied network models, as well as to the presentation of the corresponding neural mass models. The methods 
employed to characterize the linear stability of the stationary solutions for a single and two coupled populations are reported in sub-section II.D. In sub-section III.A the regions
of existence of population spikes and bursts are identified for a single population with SFA together with the bifurcation diagrams displaying the possible collective dynamical regimes.
The analysis is then extended to two symmetrically coupled excitatory or inhibitory populations without SFA in sub-section III.B and with SFA in III.C.
The relevance and influence of SFA for $\theta$-$\gamma$ CFC for two symmetrically coupled populations is examined
in Section IV. Finally a summary and a brief discussion of the results is reported in Section V. 
Appendix A summarizes the methods employed to study the linear stability of the stationary solutions for a single population.

\section{Model and Methods}

\subsection{Quadratic Integrate and Fire (QIF) Neuron}

As single neuron model we consider the QIF Neuron,
which represents the normal form of Hodgkin class I excitable membranes \cite{ermentrout1986} and it
allows for exact analytic treatements of network dynamics at the mean field level \cite{montbrio2015}.
The membrane potential dynamical evolution for an isolated QIF neuron is given by
\begin{equation}
 \tau \dot{V} (t) = V^2(t) + \eta
\end{equation}
where $\tau= 10$ ms is the membrane time constant and $\eta$ is the excitability of the neuron.

The QIF neuron exhibits two possible dynamics depending on the sign of $\eta$. 
For negative $\eta$, the neuron is excitable and for any initial condition 
$V(0) < \sqrt{-\eta}$, it reaches asymptotically the resting value $-\sqrt{-\eta}$. 
However, for initial values larger than the excitability threshold, $V(0) >\sqrt{-\eta}$, 
the membrane potential grows unbounded and a reset mechanism has to be introduced together
with a formal spike emission to mimic the spiking behaviour of a neuron. 
As a matter of fact, whenever $V(t)$ reaches a threshold value $V_{th}$, the neuron delivers a formal spike and its
membrane voltage is reset to $V_r$; for the QIF neuron $V_{th}=-V_r=\infty$.
In other words the QIF neuron emits a spike at time $t_k$ 
whenever $V(t_k^-) \to \infty$, and it is instantaneously reset to $V(t_k^+) \to - \infty$.
For positive $\eta$, the neuron is supra-threshold and it delivers a regular train of spikes with frequency $\sqrt{\eta}/\pi$.

\subsection{Network models of QIF neurons}

We consider a heterogeneous network of $N$ fully coupled QIF neurons with spike-frequency adaptation (SFA).
The membrane potential dynamics of QIF neurons can be written as
\begin{subequations}
\label{eq:QIFNEURALNETWORK}
\begin{eqnarray}
   \tau \dot{V}_i (t) &=& V^2_i(t) + \eta_i + JS(t) - A_i(t) \\
 \tau_A \dot{A}_i (t)&=&-A_i(t)+ {\alpha}\sum_{m | t_m^i < t} \delta(t-t^{i}_m)  
\\
&& \qquad \qquad  \qquad
\,\,\,\,\,\,
i=1,\dots,N \nonumber \\
\tau_S\dot{S}(t) &=&-S(t) + \frac{1}{N}\sum_{j =1}^N \sum_{k | t_k^j < t} \delta(t-t^{j}_k)
       \quad ,    
\end{eqnarray}
\end{subequations}
where the network dynamics is given by the evolution of $2N +1$ degrees of freedom. Here, $\eta_i$ is the excitability of the $i$-th neuron and $J$ is the synaptic 
strength which is assumed to be identical for each synapse. The sign of $J$ determines if the pre-synaptic neuron is excitatory ($J>0$) or inhibitory ($J < 0$). Moreover, $S(t)$ is the global synaptic current accounting for all the previously emitted spikes in the network,
where $t^{j}_k < t$ is the spike time emission of the $k$-th spike delivered by neuron $j$.
We assumed exponentially decaying PSPs with decay rate $\tau_S$, therefore $S(t)$ is simply the linear super-position of all
the PSPs emitted at previous times in the whole network. Since we have considered a fully coupled network, 
 $S(t)$ is the same for each neuron. The adaptation variable $A_i(t)$ accounts for the decrease in the excitability due to the activity of neuron $i$. Each time the neuron emits a spike at time $t^{i}_k$, the variable $A_i$ is increased by a quantity $\alpha$ and the effect of the spikes is forgotten exponentially with a decay constant
$\tau_A$. We termed this version of the network model $\mu$-SFA, since it accounts for the adaptability at a microscopic level.

However, as shown in \cite{taher2020}, by assuming that the spike trains received by each neuron
have the same statistical properties of the spike train emitted by a single neuron
(apart obvious rescaling related to the size), the SFA can be included in the model
also in a mesoscopic way. In this case the evolution equations for the membrane potentials read as
\bes  \label{eq:macroQIF}
\begin{eqnarray}
   &&\tau \dot{V}_i (t) = V^2_i(t) + \eta_i + JS(t) - A(t) \\
\nonumber && \qquad \qquad i=1,\dots,N \\
&&\tau_A \dot{A} (t) = -A(t)+ \frac{\alpha}{N} \sum_{j=1}^N \sum_{k | t_k^j < t} \delta(t-t^{j}_k)   
\\
&&\tau_S\dot{S}(t) = -S(t) + \frac{1}{N}\sum_{j =1}^N \sum_{k | t_k^j < t } \delta(t-t^{j}_k)
       \quad ,  
\end{eqnarray}
\ees
where the number of ODEs describing the network dynamics is now reduced to $N+2$ and the SFA dynamics is common to all
neurons and driven by the population firing rate 
\begin{equation}
r(t) = \frac{1}{N}\sum_{j =1}^N \sum_{k | t_k^j < t } \delta(t-t^{j}_k) \quad .
\label{eq:rate}
\end{equation}
In the following we will denote this network model as $m$-SFA. The treatment of the $N$ adaptability variables
$\{ A_i \}$ in terms of a single mesoscopic one is clearly justified i) for relatively narrow distributions of the excitabilities
with respect to the median excitability value \cite{taher2021,gast2021}  and ii) for sufficiently long adaptive time scale $\tau_A >> \tau_S$.
While the former assumption implies limiting the variability of the firing rates of the single neurons, the latter allows us to 
neglect the modulations of the firing rates on synaptic time scales \cite{gast2020mean}.

In large part of the paper, we will consider adimensional time units,
therefore the variables entering in \eqref{eq:QIFNEURALNETWORK} and \eqref{eq:macroQIF}
will be rescaled as follows
\begin{equation}
\tilde{t} = \frac{t}{\tau} \; \tilde{S} = \tau S, \; \tilde{A_i} = \tau A_i, \; \tilde{A} = \tau A
\end{equation}
and the time scales as
\begin{equation}
\tau_s = \frac{\tau_S}{\tau}, \, \tau_a = \frac{\tau_A}{\tau} \quad .
\end{equation}
 
Only in the final Secs. IV and V, we will come back to dimensional time units in order to
make easier the comparison with experimental findings.
To simplify the notation and without lack of clarity we will omit in the 
following the $(\enskip\tilde{}\enskip)$ symbol on the adimensional variables and parameters.

\subsection{Neural mass models}

\subsubsection{Single neural population}

As shown in \cite{montbrio2015}, a neural mass model describing the macroscopic evolution of a fully coupled heterogeneous QIF spiking
network with instantaneous synapses can be derived analytically, by assuming that 
the excitabilities $\left\lbrace \eta_i \right\rbrace$ follow a Lorentzian distribution
\begin{equation}
\label{eq:dist_eta}
g(\eta) = \frac{1}{\pi}\frac{\Delta}{(\eta - \bar{\eta})^2 + \Delta^2} ,
\end{equation}
where $\bar{\eta}$ is the median value of the distribution and $\Delta$ is the half-width at half-maximum (HWHM),
accounting for the dispersion of the distribution. The derivation is possible for the QIF neuronal model, since its dynamical evolution can be rewritten in terms of purely sinusoidal functions of 
a phase variable.  This allows us to apply the Ott-Antonsen Ansatz, introduced 
for  phase oscillator networks \citep{ott2008}, in the context of spiking neural networks \cite{luke2013,coombes2019}.
In particular, the analytic derivation reported in \cite{montbrio2015} allows us
to rewrite the network dynamics in terms of only two collective variables: the population firing rate $r(t)$ and
the mean membrane potential $v(t) = \sum_{i=1}^N V_i(t) /N $.  The neural mass thus introduced can be extended to include finite synaptic decays; for 
exponentially decaying synapses it takes the following form \cite{devalle2017firing}
\bes \label{eq:mean_field}
\begin{eqnarray} 
\dot{r} &=& \frac{\Delta}{\pi} + 2rv \\
\dot{v} &=& v^2 + \bar{\eta} - (\pi r)^2 + Js\\
\tau_s \dot{s} &=& -s + r \enskip ,
\end{eqnarray}
\ees
where a third variable in now present, $s(t)$, representing the global synaptic field.

The inclusion of SFA in the neural mass model Eqs. \eqref{eq:mean_field} is straightforward when considering the $m$-SFA,
since, in this context, the adaptability is described by a collective variable. This finally leads
to the following four dimensional mesoscopic model for a single QIF population
\bes  \label{eq:mean_field2}
\begin{eqnarray}
\label{r}
 \dot{r} &=& \frac{\Delta}{\pi} + 2rv \\
 \label{v}
 \dot{v} &=& v^2 + \bar{\eta} - (\pi r)^2 + Js - A \\
 \label{s}
 \tau_s \dot{s} &=& -s + r \\
 \label{a}
\tau_a \dot{A} &=& -A +  \alpha r \quad ,
\end{eqnarray}
\ees
which represents the exact mean field formulation of the QIF spiking network with
$m$-SFA, described by the $N+2$ set of ODEs \eqref{eq:macroQIF}, in the limit $N \to \infty$.
However, as we will show thereafter, it can also capture 
the dynamics of the network with $\mu$-SFA, described by the set of ODEs \eqref{eq:QIFNEURALNETWORK} and characterized by $N^2 +1$ variables,
for limited neural hetereogeneity (i.e. sufficiently small $\Delta$) and sufficiently long adaptive time scales $\tau_a$.


\subsubsection{Two symmetrically coupled neural population}
 
The dynamics of two symmetrically coupled identical neural masses with adaptation is described by the 8-dim system of ODEs
\bes     \label{eq:two_pop_adapt}
\begin{eqnarray}
&&        \dot{r}_{1,2} = \frac{\Delta}{\pi} + 2r_{1,2}v_{1,2} \\ 
&&         \dot{v}_{1,2} = v_{1,2}^2 + \bar{\eta} - (\pi r_{1,2})^2 + J_{s}s_{1,2}+ J_{c}s_{2,1} - A_{1,2}\\
&&         \tau_{s}\dot{s}_{1,2} = -s_{1,2} + r_{1,2}\\
&&			     \tau_{a}\dot{A}_{1,2} = -A_{1,2} + \alpha r_{1,2}	   \quad ;      
\end{eqnarray} 
\ees
where $J_s$ and $J_c$ represent the self- and cross-coupling, respectively.

On the basis of Eqs. \eqref{eq:two_pop_adapt}, one cannot distinguish between the two populations,
therefore these equations are invariant under the permutation of the variables
$(r_1, v_1, s_1, A_1, r_2, v_2, s_2, A_2) \longrightarrow (r_2, v_2, s_2, A_2, r_1, v_1, s_1, A_1) $
and they admit the existence of entirely symmetric solutions $(r_1, v_1, s_1, A_1) \equiv (r_2, v_2, s_2, A_2)$.

By following \cite{ratas2017symmetry}, we analyse the stability of symmetric solutions by 
transforming the original set of variables in the following ones 
\begin{equation}
\begin{pmatrix}
r_{l,t} \\
v_{l,t} \\
s_{l,t} \\
A_{l,t}
\end{pmatrix}  = \frac{1}{2} \left\lbrace \begin{pmatrix} r_2 \\ v_2 \\ s_2 \\ A_2 \end{pmatrix}  \pm \begin{pmatrix} r_1 \\ v_1 \\ s_1 \\ A_1 \end{pmatrix} \right\rbrace 
\label{eq:transv_variables}
\end{equation}
where we refer to  $(r_l, v_l, s_l,A_l)$ and $(r_t, v_t, s_t,A_t)$ as the longitudinal and transverse set of coordinates, 
respectively. 

In this new set of coordinates, the trajectories of the symmetric solutions live in the invariant 
subspace $(r_l, v_l, s_l,  A_l, r_t \equiv 0, v_t \equiv 0, s_t \equiv 0, A_t \equiv 0)$, with the longitudinal variables satisfying the following set of ODEs 
\bes \label{eq:long_sym_sys}
\begin{eqnarray}
 	\Dot{r_l} &=& \frac{\Delta}{\pi} + 2 r_l v_l \\
   \Dot{v_l} &=& v_l^2 + 1 - (\pi r_l)^2 + (J_{s} + J_{c})s_l -A_l \\
   \tau_s \Dot{s_l} &=& s_l - r_l \\
		    \tau_a \Dot{A_l} &=& A_l - \alpha r_l  \enskip .    
\end{eqnarray}
\ees

\subsection{Linear Stability Analysis}

\subsubsection{Stationary solutions for a single population}

By following \cite{devalle2017firing}, we explore the stability of the stationary solutions of the single QIF population, corresponding to fixed point solutions $(r_0,v_0,s_0,A_0)$ in the neural mass formulation \eqref{eq:mean_field2}. These fixed point solutions 
are given by ${s}_0= { r}_0$, ${A}_0=\alpha {r}_0$, together with the implicit algebraic system
\begin{eqnarray}
    \label{eq:equilibriumR} {v}_0 &=& - \frac{\Delta}{2 \pi {r}_0} \\
    \label{eq:equilibriumV} 0 &=& {v}_0^2 + \bar{\eta} + J {r}_0 - \pi^2 {r}_0^2 - \alpha {r}_0 .
\end{eqnarray}
Notice that the whole equilibrium solution can be parametrized in terms of ${r}_0$. 

The linear stability analysis of the fixed point can be performed by estimating the eigenvalues 
$\lambda$ of the associated Jacobian matrix
\begin{equation}
\label{eq:jacobian_one_pop}
        			\mathbf{H} = \begin{pmatrix} 
                        2 {v}_0 & 2 {r}_0 & 0 & 0\\
                        -2 \pi ^2 {r}_0 & 2 {v}_0 & J & -1 \\
                        \frac{1}{\tau_s} & 0 & -\frac{1}{\tau_s} & 0 \\
                        \frac{\alpha}{\tau_a} & 0 & 0 & -\frac{1}{\tau_a} \enskip .\\
                     \end{pmatrix}
\end{equation}
The eigenvalues are the solutions of the corresponding  characteristic polynomial  $p(\lambda)$.
In particular, assuming $\lambda = i \Omega$ leads to the Hopf 
bifurcation curves, while setting $\lambda = 0$ leads to saddle-node and pitchfork bifurcations. 
This analysis is performed in details in Appendix A.

\subsubsection{Stationary solutions for two coupled populations}

The stability of the symmetric solutions of Eqs. \eqref{eq:long_sym_sys} can be examined in the longitudinal or transverse 
manifold by employing the transformation of variables reported in \eqref{eq:transv_variables}.
In particular, as seen in many contexts ranging from chaotic maps \cite{pikovsky1991}
to chimera states \cite{olmi2015}, the longitudinal instabilities preserve the symmetric nature of the solutions,
while the transverse ones are responsible for symmetry breaking.

Here, we will focus on the linear stability of symmetric stationary solutions, corresponding
to fixed points $({\bar r}_l,{\bar v}_l,{\bar s}_l,{\bar A}_l)$ of Eqs. \eqref{eq:long_sym_sys}. The linear and transverse 
stability analysis can be performed by considering the linearized evolution around the fixed points
given by 
\begin{equation}
           \begin{pmatrix}
           \delta \Dot{r}_{l,t} \\
           \delta \Dot{v}_{l,t} \\
           \delta \Dot{s}_{l,t} \\
           \delta \Dot{A}_{l,t} \\           
         \end{pmatrix} = \mathbf{H}_{l,t} \begin{pmatrix}
           \delta r_{l,t} \\
           \delta v_{l,t} \\
           \delta s_{l,t} \\
           \delta A_{l,t} \\           
         \end{pmatrix}
\end{equation}
with
\begin{equation}
\label{eq:mat_longitudinal}
    \mathbf{H}_{l,t} =            \begin{pmatrix}
           2 {\tilde v}_{l,t} & 2 {\tilde r}_{l,t} & 0 & 0\\
           -2 \pi^2 {\tilde r}_{l,t}  & 2 {\tilde v}_{l,t} &  J_{s} \pm J_{c} & -\frac{1}{\tau_a} \\
            \frac{1}{\tau_s} & 0 & - \frac{1}{\tau_s} & 0\\
           \frac{\alpha}{\tau_a}  & 0 & 0 & -\frac{1}{\tau_a} 
         \end{pmatrix}.
\end{equation}
The longitudinal and transverse stability of the fixed points can be analysed by considering the eigenvalue
problems associated to $\mathbf{H}_{l}$ and  $\mathbf{H}_{t}$, respectively.
In particular, we notice that the stability can be lost in two different ways: 
i) one eigenvalue becomes exactly equal to $0$ (similarly to a limit point); 
ii) the real part of two complex conjugates eigenvalues becomes exactly zero 
(Hopf bifurcation).

Therefore we have a total of 4 possible changes of
stability of the symmetric stationary solutions giving rise to the following bifurcations
in the complete eight dimensional phase space:

\begin{itemize}

\item{\textit{Longitudinal Limit Point bifurcation (LLP):}} Saddle-node bifurcation in the original system;

\item{\textit{Longitudinal Hopf bifurcation (LH):}} Hopf bifurcation generating symmetric oscillations;

\item{\textit{Transverse Limit Point bifurcation (TLP):}} Branch-Point or pitchfork bifurcation in the original system
giving rise to two different equilibria for the two populations;
 
\item{\textit{Transverse Hopf bifurcation (TH):}} Hopf bifurcation in the original system generating asymmetric oscillations for the two populations, which can differ simply in their phase or even in amplitude and phase.

\end{itemize}

The transverse and longitudinal stability analysis of the symmetric solutions is verified and
complemeted by employing standard continuation packages (such
as MATCONT \cite{dhooge2003matcont}) over the dynamics of the complete system \eqref{eq:two_pop_adapt},

\begin{figure}
\includegraphics[width=0.95\linewidth]{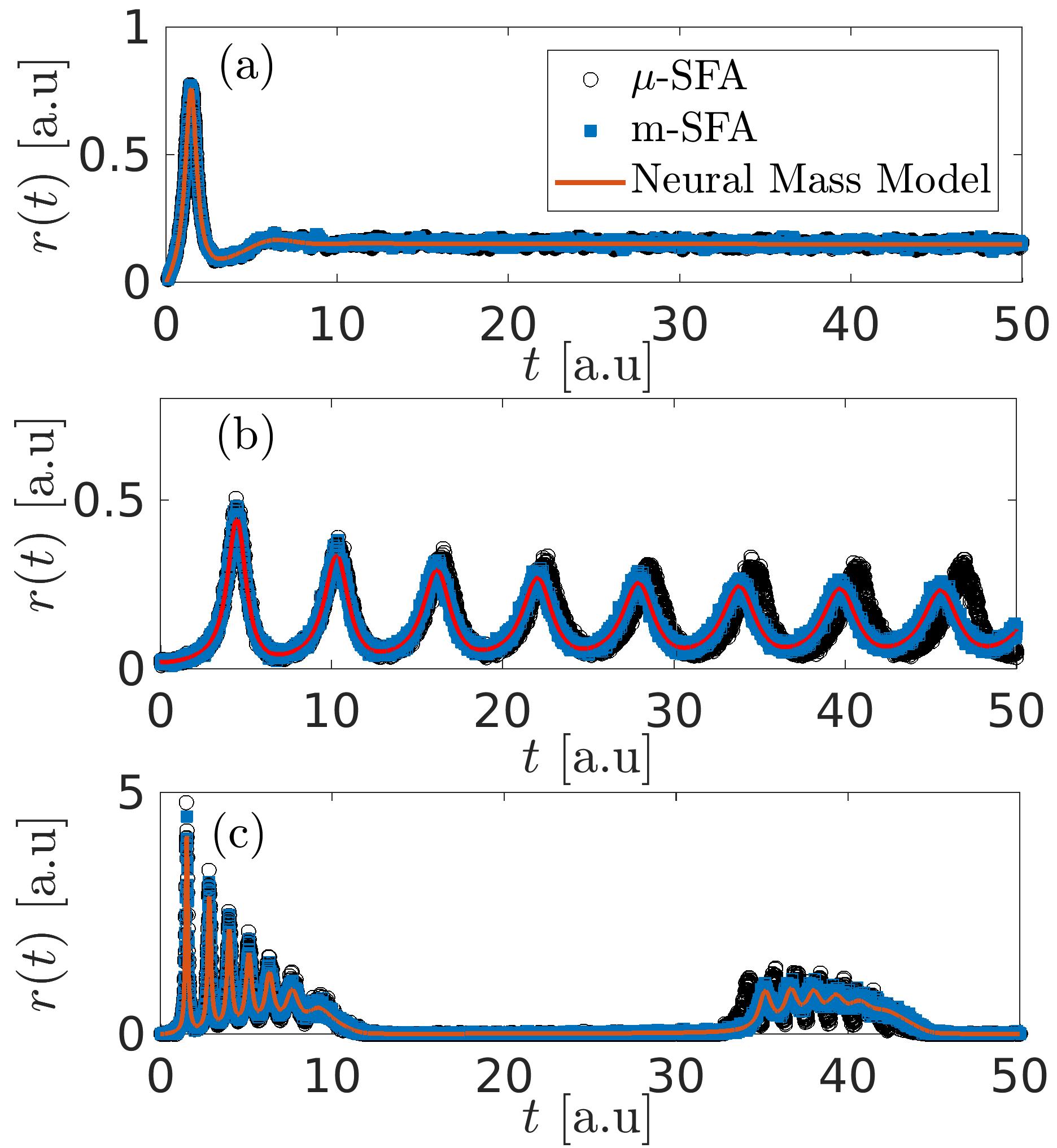}
\caption{Comparison of population firing rate $r(t)$ as obtained by the network models with $\mu$-SFA \eqref{eq:QIFNEURALNETWORK} (black circles) and with $m$-SFA  \eqref{eq:macroQIF} (blue squares),
with the evolution of the neural mass models  \eqref{eq:mean_field} (red line) for different macroscopic behaviours :
a) stationary state (parameters $J = -6$, $\Delta = 0.5$, $\alpha = 1$); b)
population spiking ($J = -6$, $\Delta = 0.12$, $\alpha = 1$); and c) population bursting ($J = 15$, $\Delta = 0.12$, $\alpha = 20$). Other parameters are $\tau_s = 1.5$, $\tau_a = 10$ and $\bar{\eta} = 1$. 
For the network simulations the number of considered QIF neurons is $N=10,0000$ (panel A) or 
$N=100,000$ (panels B and C). For all comparisons the neural mass model has been integrated starting 
from initial values of $r,v,A$ and $s$ as obtained from the microscopic state of the considered networks.}
\label{fig:dynamics_simple}
\end{figure}

\section{Results}

The analysis reported in this paper will be focused on neural mass models, however, before reporting
the corresponding results, let us compare in a few cases the simulations done separately for the network and the neural masses.
As previously stated, the neural mass model \eqref{eq:mean_field} reproduces exactly the population dynamics 
of a network of QIF spiking neurons with $m$-SFA \eqref{eq:macroQIF} for a sufficienltly large number of neurons.
Indeed, as shown in Fig \ref{fig:dynamics_simple}, the agreement between the neural mass (red solid line) and the network simulations (blue squares) is very good already for $N > 10,000$ for different macroscopic regimes observable in  purely inhibitory or excitatory populations.
For what concerns the $\mu$-SFA network model \eqref{eq:QIFNEURALNETWORK} (black circles), while in the asynchronous regime (stationary state) the
agreement with the other two models is noticeable (see Fig. \ref{fig:dynamics_simple} (a)), in presence of collective oscillations there is a clear dephasing between neural mass and $\mu$-SFA dynamics at sufficiently long times (see Fig. \ref{fig:dynamics_simple} (b,c)).
In general, the $\mu$-SFA networks will display the same macroscopic states and bifurcation structure as the neural mass models, even if definitely greater discrepancies are expected for low median excitability with
a high level of heterogeneity or for  faster adaptation time scales \cite{gast2020mean,taher2021,gast2021}.

\subsection{Single population with SFA}
\label{sec:single_pop}

\begin{figure}
\includegraphics[width=0.95\linewidth]{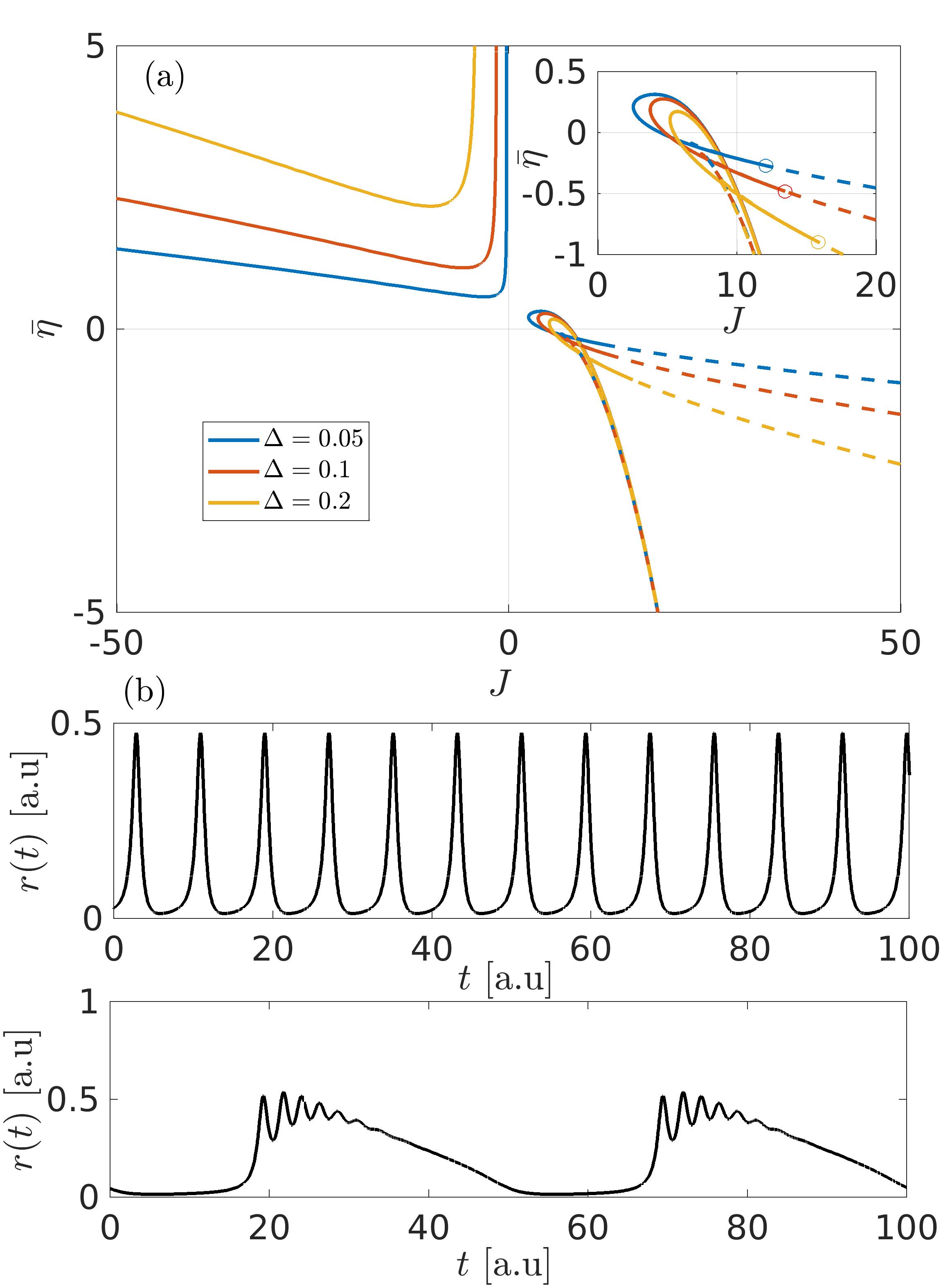}
\caption{{\bf Single neural mass with SFA} a) Hopf boundaries in the $(J,\bar{\eta})$ space at different values of $\Delta$. Parameters used here $\tau_s = 2$, $\tau_a = 10$ and $\alpha = 5$. Solid lines refer to Hopf curves while dashed lines refer to saddle-node curves. At $J>0$ the Hopf curves meet with the saddle node curves at Bogdanov Takens points (circle), as shown in the inset. b) Oscillations generated within the Hopf region in the inhibitory case with $\eta = 1$ and $J = -7$ (top) and the excitatory case with $\eta = 0.1$ and $J = 7$ (bottom). In both cases
$\Delta = 0.05$ was used.}
\label{fig:Hopf_boundary_J_eta}
\end{figure}

In this subsection we analyse the relevance of SFA for the emergence of collective dynamics. In particular,
we consider a single heterogeneous population
of QIF neurons with SFA. All the results here reported are based on the linear stability analysis of the stationary solutions
of the neural mass models, as discussed in subsection II.A.1 and Appendix A.

In the absence of adaptation, the excitatory population with exponentially decaying synapses displays only fixed point solutions (namely, foci): for sufficiently negative (positive) ${\bar \eta}$, a single stable fixed point is present corresponding to a low (high) firing rate solution, while just below  ${\bar \eta} = 0 $, the two solutions coexist. For exponentially decaying synapses
both fixed points are foci, characterized by one real and two complex conjugate eigenvalues, at variance with what reported  in the absence of a synaptic time scale \cite{montbrio2015}, where the low firing solution is a node, not a focus.
On the other hand, in the inhibitory case, one observes tonic population spiking in a limited range of synaptic time scales $\tau_s \simeq 0.1 - 100$ and for not too wide heterogeneity distributions and synaptic coupling strenghts $|J|$ \cite{devalle2017firing,ceni2020}. In this case the period of the COs is essentially controlled by the synaptic time
scale, since the oscillations emerge due to the inhibitory action that desynchronizes the spiking activity \cite{devalle2017firing}. 

In the presence of SFA, COs are observable both in purely excitatory and inhibitory populations,
as shown in Fig. \ref{fig:Hopf_boundary_J_eta}(a), where the Hopf bifurcation lines are reported (as solid lines) in the $(J,\bar{\eta})$ plane at different values of the HWHM of the heterogeneity $\Delta$. In the inhibitory case, for the present choice of synaptic and adaptation time scales, COs emerge for any value of $|J|$, for sufficiently large excitatory drive  ${\bar \eta}$, i.e. above the Hopf bifurcation curves shown in Fig. \ref{fig:Hopf_boundary_J_eta}(a) for $J <0$.
For increasing $\Delta$, the Hopf curve moves at higher values of ${\bar \eta}$, thus indicating that a higher excitatory drive is required to overcome the heterogeneity present in the population. In the excitatory case, stable COs are instead limited to a small portion of the plane in proximity of $\bar{\eta} \approx 0$ and $J \simeq 5-10$, as shown the inset of Fig. \ref{fig:Hopf_boundary_J_eta}(a). The COs are stable within the closed loop in the $(J,\bar{\eta})$  plane, while, outside the loop, the macroscopic limit cycles are unstable and eventually both branches of the Hopf curve end at a Bogdanov Takens (BT) bifurcation point of codimension two (circles in the inset) \cite{bogdanov}. As expected, at this point, saddle-node bifurcation curves of fixed points (dashed lines) emerge. 

\begin{figure}
\includegraphics[width=0.95\linewidth]{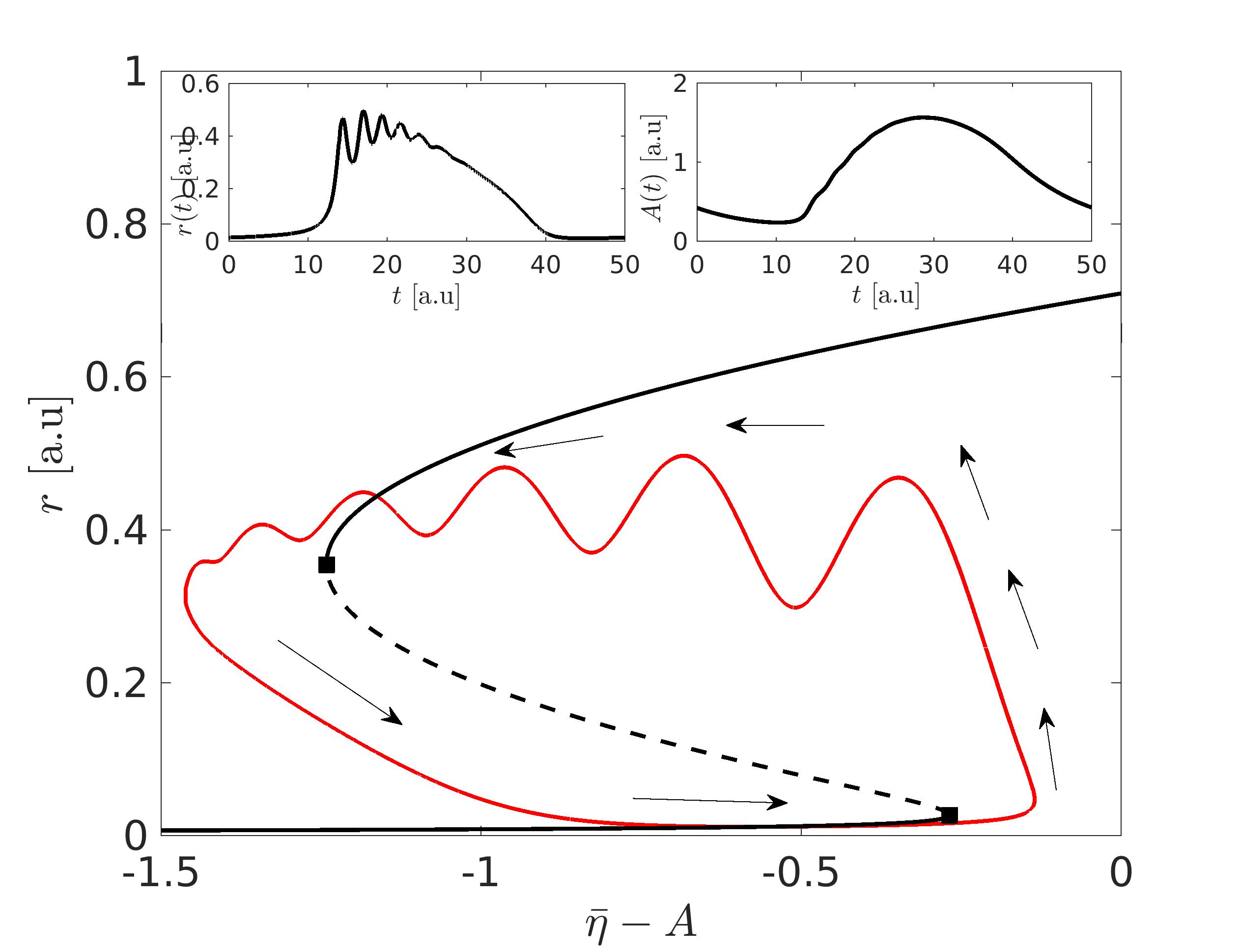}
\caption{{\bf Single neural mass with SFA} The population firing rate $r$ versus ${\bar \eta} - A$ for the population burst solution diplayed in \ref{fig:Hopf_boundary_J_eta} (b) (bottom). The solid (dashed) black lines refer to stable (unstable) fixed points of the fast
sub-sytem (\ref{r},\ref{v},\ref{s}). The filled black squares denote the saddle-node bifurcations in the fast sub-system.
The red solid line is the bursting solution of the complete system \eqref{eq:mean_field2}, where
the black arrows indicate the time evolution along such a solution.
In the upper insets are shown the population firing rate $r(t)$ and the adaptation variable $A(t)$ as a function of time during the evolution of the population burst. Parameters as in
\ref{fig:Hopf_boundary_J_eta}.
}
\label{fig:Burst_Explained}
\end{figure}

As it can be seen in Fig. \ref{fig:Hopf_boundary_J_eta}(b), COs generated via inhibition (top) are population spikes characterized by an unique oscillation period, whose value is controlled by $\alpha$ and which can range form $\tau_s$ to $\tau_a$. On the other hand excitatory mediated COs (bottom) are characterized by two frequencies: a slow one (definitely longer than $\tau_a$), responsible for the alternation of silence and high activity in the population and a fast one (of the order of $\tau_s$), controlling the fast damped oscillations developing on top of the slow carrier. These kind of COs have been termed emergent bursting or population bursting, in analogy with the bursting activity of the single neurons \cite{gast2020mean,chen_adaptation}. To better understand the origin of these population bursts we have performed a standard slow-fast analysis, by considering as fast sub-system the one with constant adaptation $A$. In particular, we analyse the possible solutions
emerging in the fast sub-system (\ref{r}, \ref{v}, \ref{s}) as a function of the control parameter $\bar{\eta} - A$.
The solutions are stable (unstable) fixed points reported as solid (dashed) black lines in  Fig. \ref{fig:Burst_Explained} and they correspond to a low and a high firing solution 
that can coexist in a certain interval of the parameter values $\bar{\eta} - A$.  Together with the fast sub-system solutions, we display also
the population burst (solid red line) obtained for the full equation system \eqref{eq:mean_field2} as $r(t)=r(\bar{\eta}-A(t))$.
As shown in the upper insets of Fig. \ref{fig:Burst_Explained}, the variable $A(t)$ displays regular periodic oscillations with period $\tau_a$, each complete oscillation
corresponding to a population burst in the variable $r(t)$.

By examining the evolution of the bursting solution in Fig. \ref{fig:Burst_Explained},
we observe that the orbit is initially lying on the lower stable solution branch
of the fast sub-system. However, during the burst evolution (indicated by the black arrows) the variable  $A(t)$ decreases.
This decrease finally leads the orbit beyond the region of existence of the lower branch, where only the higher branch 
state is present and stable. Afterwards, the orbit is attracted towards the focus in the upper branch and approaches it by displaying damped
oscillations with periods of the order of $\simeq 1.1 - 2.8$. In the meantime the variable $A$ is growing. 
The growth of $A$ leads the burst orbit beyond the region of existence of the upper state and, consequently, the orbit is attracted back towards the only stable solution, i.e. the lower state,
and the orbit evolution repeats. However, it still remains an aspect to be clarified regarding this slow-fast analysis: why the orbit approaches the
upper state by displaying damped oscillations, while it approaches the lower state without such oscillations, despite
both solutions are foci of the fast sub-system. This is due to the fact that the period of the damped oscillations
in the lower branch is definitely larger than the characteristic period of the population bursts
$t_a$ (namely, $\simeq 12-40$),  and therefore oscillations cannot be observed on this time scale.

\begin{figure}
\includegraphics[width=0.95\linewidth]{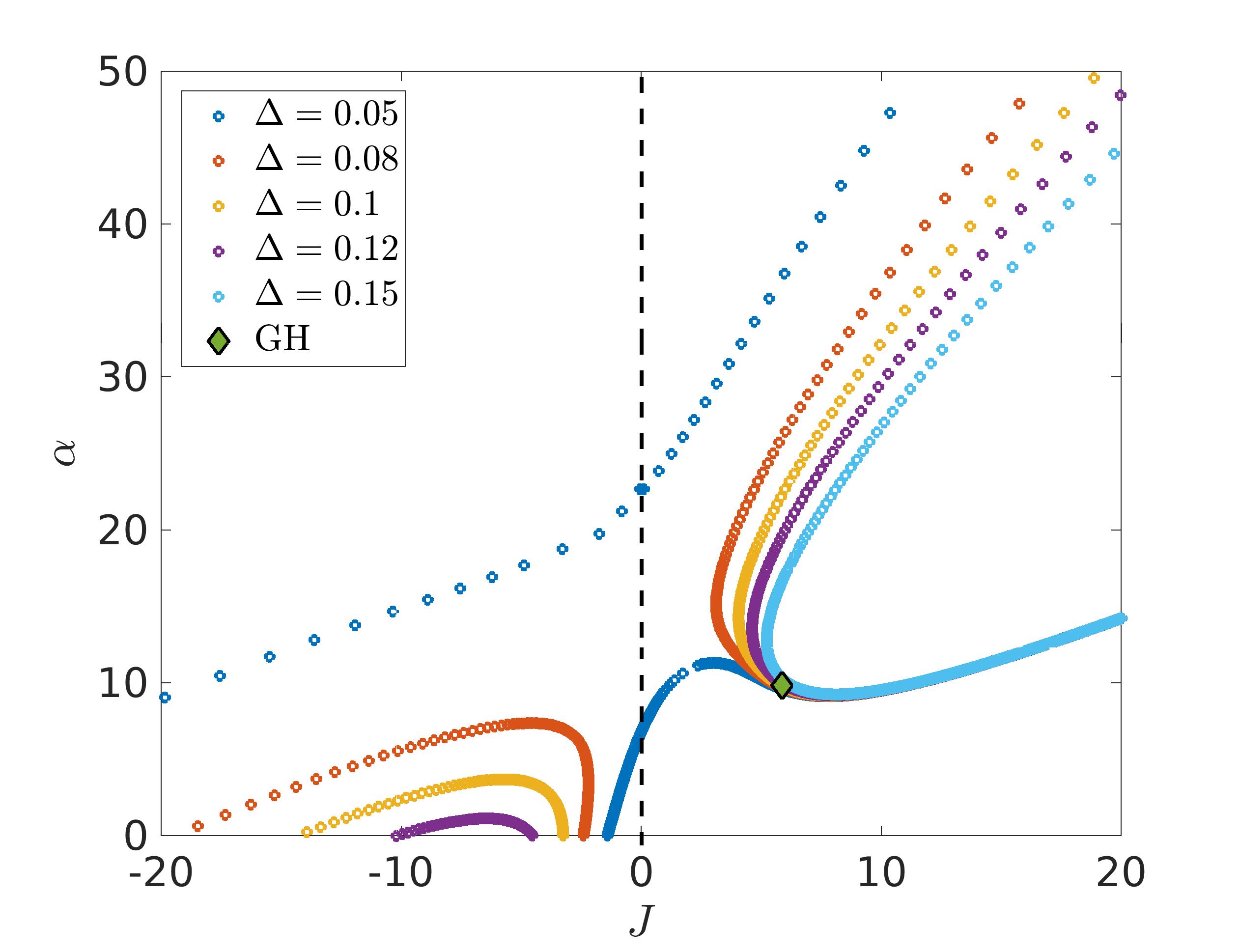}
\caption{{\bf Single neural mass with SFA} Hopf boundaries in the $(J,\alpha)$ space at different  values of $\Delta$. Parameters used here: $\tau_s = 2$, $\tau_a = 10$ and $\bar{\eta} = 1$. Dashed vertical line separates the excitatory regime from the inhibitory one.
The symbol referes to the Generalized Hopf (GH) bifurcation point for
$\Delta = 0.1$ located at $(J,\alpha) = (5.86,9.81)$.}
\label{fig:Hopf_boundary_J_alfa}
\end{figure}

The effect of the adaptation on the population dynamics can be better appreciated by considering the bifurcation diagram in the $(J,\alpha)$ plane for different $\Delta$-values, as shown in Fig. \ref{fig:Hopf_boundary_J_alfa}. As already mentioned, in the excitatory case no COs are observable in the system in the absence of adaptation. Instead, in the presence of SFA there is
a wide region where population bursts can be observed for sufficiently large $\alpha$ and for any $\Delta$: the region inside 
each Hopf bifurcation curve, identified by circles of different colors for each $\Delta$-value in Fig. \ref{fig:Hopf_boundary_J_alfa}.
In this case, on the lower Hopf bifurcation curve, a Generalized Hopf (GH) point
is observable, distinguishing super-critical (at smaller $J$) from sub-critical 
Hopf-bifurcations (at larger $J$); an example is shown for $\Delta=0.1$ in Fig. \ref{fig:Hopf_boundary_J_alfa}.
Just above the sub-critical line one has a region where population bursts and stationary foci coexist, similarly to what reported in \cite{gast2020mean} for an adaptation current with an $\alpha$-kernel and instantaneous synapses.

For the inhibitory case an interval of synaptic strength $J$ where COs are present can be identified even at $\alpha=0$,
for sufficiently small $\Delta	\lesssim 0.13$ : it is the interval between the crossing of the Hopf curves with the zero axis in Fig. \ref{fig:Hopf_boundary_J_alfa}. 
The presence of adaptation shrinks the interval where COs can be observed, having in this case an opposite effect with respect to what seen for an excitatory population.  
This is true for sufficiently large heterogeneity $\Delta > \Delta_c \simeq 0.063271$, while, below the critical value $\Delta_c$, the Hopf curves for excitatory and inhibitory populations merge together,
giving rise to a wide region where COs are observable (see the blue circles in Fig. \ref{fig:Hopf_boundary_J_alfa} delimiting the internal region of existence of COs for $\Delta=0.05$).

In summary, the adaptation has a twofold effect of promoting COs in excitatory neural masses and preventing their emergence in inhibitory ones. This can be explained by the fact that
in the excitatory case the only source of inhibition arises from SFA. As a consequence, we can expect that the adaptation time scale plays a
role somehow similar to that of the synaptic time scale in purely inhibitory networks for the emergence of COs \cite{whittington1995}. In the inhibitory case the effect of adaptation is to increase the inhibitory effect and to slow down the inhibitory response, thus, for sufficiently large $\alpha$, the COs can eventually disappear 
for a mechanism similar to that reported in \cite{devalle2017firing,ceni2020} in the absence of adaptation.

\subsection{Two coupled populations without SFA}
\label{sec:alpha0}

In this sub-section we analyse the dynamical regimes emerging in two symmetrically coupled neural  
mass models without adaptation, corresponding to Eqs. \eqref{eq:two_pop_adapt} with $\alpha=0$.
In particular, we vary the cross-coupling and the median excitability as control parameters, thus evaluating the bifurcation diagram in the bidimensional plane $(J_c,{\bar \eta})$.
The synaptic time scale $\tau_s$, the level of heterogeneity $\Delta$ and the self-coupling $J_s$ are kept fixed.

\subsubsection{Two inhibitory populations}

\begin{figure*}
\includegraphics[width=1\linewidth]{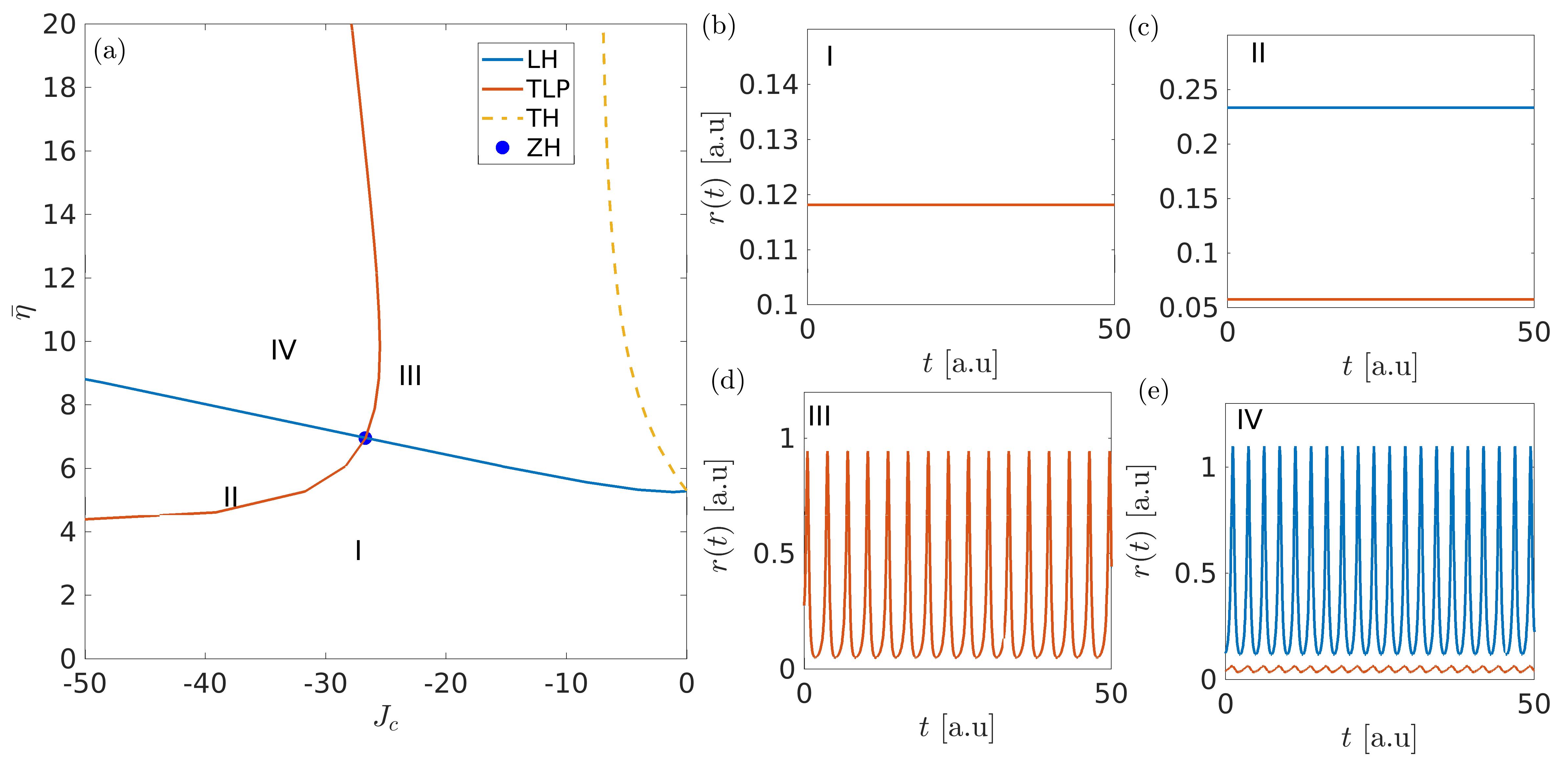}
\caption{{\bf Two symmetrically coupled inhibitory populations without adaptation} 
a) Phase diagram in the $(J_{c},\bar{\eta})$ space showing the longitudinal
(blue solid line) and transverse (red solid line) stability boundaries. Stability boundaries of stable (unstable) solutions are shown as (solid) dashed lines. Longitudinal and transverse stability boundaries meet at a Zero-Hopf point dividing 
the parameter space in four regions. A sample time trace of the firing rate of each population 
is shown in panels (b-e) for the four possible dynamical regimes : b) region I, symmetric fixed points $(J_{c},\bar{\eta})=(-25,5)$;
c) region II, asymmetric fixed points $(J_{c},\bar{\eta})=(-33,7)$;
d) region III, symmetric COs $(J_{c},\bar{\eta})=(-25,10)$;
e) region IV, asymmetric COs $(J_{c},\bar{\eta})=(-33,10)$. 
The blue (red) solid lines in (b-e) denotes the firing rate of population 1 (population 2).
Other parameters for this figure: $\tau_s = 2$, $\Delta = 0.5$, $J_{s} = -20$, $\alpha=0$.}
\label{fig:symmetry_break_alpha0_inh}
\end{figure*}

Let us first consider two inhibitory neural masses that display, when uncoupled,
a stationary solution (fixed point) for sufficiently small
${\bar \eta } < \eta_c$ , and COs for larger ${\bar \eta } >  \eta_c$.
Once symmetrically coupled, the two populations can display various dynamical regimes emerging  from the stable fixed points due to longitudinal or transverse instabilities, as shown in Fig. \ref{fig:symmetry_break_alpha0_inh}.
In the present analysis as well as in the following, we will only consider transitions associated to stable states (solid lines in Fig. \ref{fig:symmetry_break_alpha0_inh}) and we will neglect those corresponding to unstable states (dashed lines).

In particular, the longitudinal Hopf (LH) instability (blue solid line in Fig. \ref{fig:symmetry_break_alpha0_inh}(a)) divides the plane $(J_c,{\bar \eta})$ in two parts : one associated to fixed points (COs) below (above) the line. The LH line is the continuation of the critical point $\eta_c$, found at $J_c \equiv 0$, to finite cross-couplings. Another bifurcation curve corresponding to a transverse
instability is reported in Fig. \ref{fig:symmetry_break_alpha0_inh}(a). This is a TLP bifurcation line (red solid curve): the dynamical evolution of the two populations is identical (different) to the right (to the left) of this curve. The LH and TLP bifurcation lines meet at a Zero-Hopf (ZH) point, thus
creating 4 different regions with distinct types of dynamics.
For each region the evolution of the firing rate of population 1 (population 2) is shown in Fig. \ref{fig:symmetry_break_alpha0_inh} (b-e) as blue (red) solid line, and they correspond to the following regimes: 
\begin{itemize}
\item region I: the two firing rates are constant and identical (see  Fig. \ref{fig:symmetry_break_alpha0_inh} (b));
\item region II: the two firing rates are constant but with different values (see Fig. \ref{fig:symmetry_break_alpha0_inh} (c));
\item region III: the two firing rates oscillate in perfect synchrony (see Fig. \ref{fig:symmetry_break_alpha0_inh} (d));
\item region IV: the two firing rates oscillates with different amplitudes, the two populations oscillate in phase
but their activity is different (see Fig. \ref{fig:symmetry_break_alpha0_inh} (e)).
\end{itemize}
The collective periodic oscillations here reported can be all identified as population spikes \cite{varona2000}, where a large part of the neurons fires within a short time window and their periods are obviously related to the synaptic time scale $\tau_s$ \cite{devalle2017firing}.

The non identical solutions observable in Fig. \ref{fig:symmetry_break_alpha0_inh}(c) and
(e) correspond to a spontaneous symmetry breaking of the dynamical evolution
and they emerge for $|J_c| > |J_s|$. In particular,  these solutions can be identified as generalized chimera states  \cite{olmi2011, ratas2017symmetry}, where the two populations display different levels of synchronization characterized by constant (as in panel (c)) or periodically (as in panel (e))
oscillating Kuramoto order parameters \cite{kura}, as we have verified by employing the conformal transformation introduced in \cite{montbrio2015}, which relates the Kuramoto order parameter 
to the mean membrane potential and the population firing rate.

\subsubsection{Two excitatory populations}
\label{2exc}

\begin{figure*}
\includegraphics[width=1\linewidth]{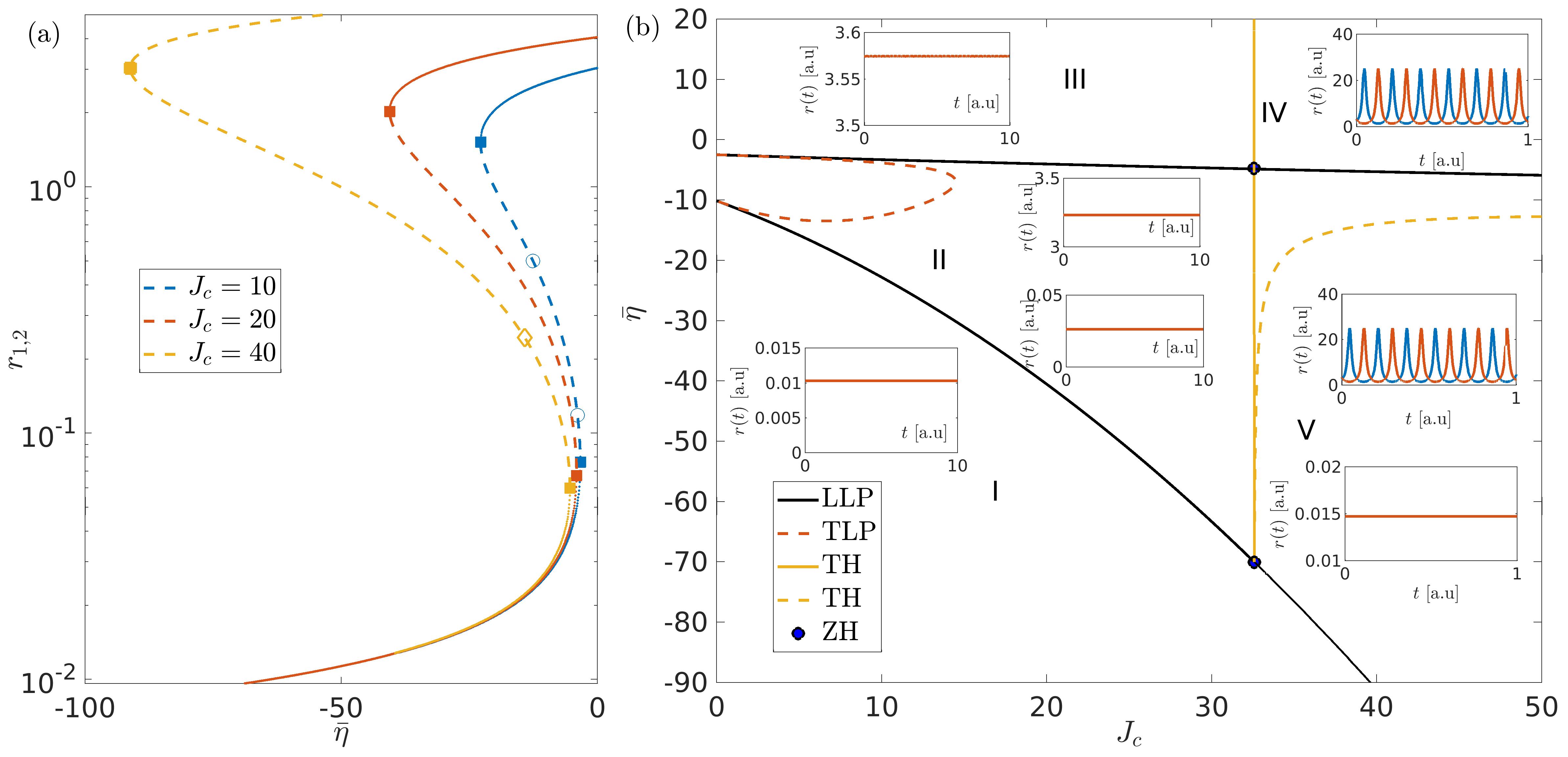}
\caption{{\bf Two symmetrically coupled excitatory populations without adaptation} 
a) Evolution of the population firing rates $r_{1,2}$ versus ${\bar \eta}$
for the stationary solution corresponding to an asynchronous regime. The fixed point solutions
are shown for three different values of the cross-coupling: namely, $J_c=10$ (blue lines), 20 (red lines) and 
40 (yellow lines). Stable (unstable) solutions are denoted by solid (dashed) lines, 
saddle-node bifurcations ae denoted by filled squares, pitchfork bifucations
by empty circles and Hopf bifurcations by empty diamonds.
b) Two dimensional phase diagram in the $(J_{c},\bar{\eta})$ plane showing the saddle node curves (LLP)
as black solid curves and the transverse Hopf (TH) boundary as yellow solid line. The dashed lines refer
to transverse instabilities connecting unstable solutions, which are therefore not observable in the
system evolution. Other parameters for this figure: $\tau_s = 2$, $\Delta = 0.5$, $J_{s} = 20$. $\alpha=0$.}
\label{fig:symmetry_break_alpha0_exc}
\end{figure*}

As a complementary analysis, we consider now two identically coupled excitatory neural masses. 
In this case, for sufficiently low values of the cross-coupling $J_c$, we observe two coexisting symmetric fixed points,
corresponding to a low and a high activity state of the neural population, analogously to what found for a single population. 
These solutions are shown in Fig. \ref{fig:symmetry_break_alpha0_exc} (a) as a function of ${\bar \eta}$ for increasing values of $J_c$. At low values (namely, $J_c=10$ and 20) we observe that the low activity state looses its stability by increasing ${\bar \eta}$ via a saddle-node bifurcation; this fixed point is connected to the high activity state via an unstable solution
that turns to be stable via a second saddle-node bifurcation. In particular, we observe a coexistence region between the two solutions delimited by the occurence of the two saddle-node bifurcations. 
At larger cross-coupling ($J_c=40$) the high activity state is instead always unstable.

The complete scenario can be better understood by considering the two dimensional phase diagram
in the plane  $(J_{c},\bar{\eta})$, shown in Fig. \ref{fig:symmetry_break_alpha0_exc} (b),
obtained by computing longitudinal and transverse instabilities for this system. Let us first focus on the part of phase diagram delimited by the vertical line $J_{c}\lesssim 32.5$ (yellow solid line)
where one has only symmetric solutions and three distinct regions separated by 
two longitudinal saddle-node bifurcation curves (LLP) are observables :
 
\begin{itemize}
\item region I: an unique stable low activity fixed point corresponding to the lower branch of equilibria in Fig.
\ref{fig:symmetry_break_alpha0_exc} (a);

\item region II: coexistence of two fixed points corresponding to high and low activity. The region inside
the red dashed curve corresponding to a TLP bifurcation line is not observable, since it is related to the 
continuation of the branch point appearing in the unstable branch in Fig. \ref{fig:symmetry_break_alpha0_exc} (a) at low $J_{c}$;

\item region III: an unique stable high activity fixed point corresponding to the stable upper branch of equilibria in
Fig. \ref{fig:symmetry_break_alpha0_exc} (a).
\end{itemize} 
At $J_{c}\approx 32.5$ a spontaneous symmetry breaking occurs in the system, due to a transverse Hopf instability (TH),
leading to the emergence of collective anti-phase oscillations from the high activity fixed point.
These COs are characterized by extremely fast periods of oscillation definitely smaller than the synaptic time constant $\tau$.
Here two other dynamical regimes are now observable:
\begin{itemize}
\item region IV: the stable high activity point is no longer stable, 
and an asymmetric solution where both populations oscillate in antiphase appears;

\item region V: below the upper saddle-node curve the asymmetric solution characterized by antiphase oscillations 
is still stable and coexists with the low activity fixed point. The lower branch of the saddle-node curve to the right of the Zero-Hopf point delimits the region where spontaneous symmetry breaking occurs. 
The dashed yellow curve in region V is the continuation of the Hopf point
reported in panel (a) for $J_{c}=40$, which leads to unstable oscillations of asymmetric nature (unstable TH curve).
\end{itemize}

\subsection{Two coupled populations with SFA}
\label{sec:alpha}

We now investigate the collective dynamics emerging in a system of two symmetrically coupled neural masses 
with spike frequency adaptation. We first analyze two coupled inhibitory populations and  then
two coupled excitatory ones.

\begin{figure*}
\includegraphics[width=0.95\linewidth]{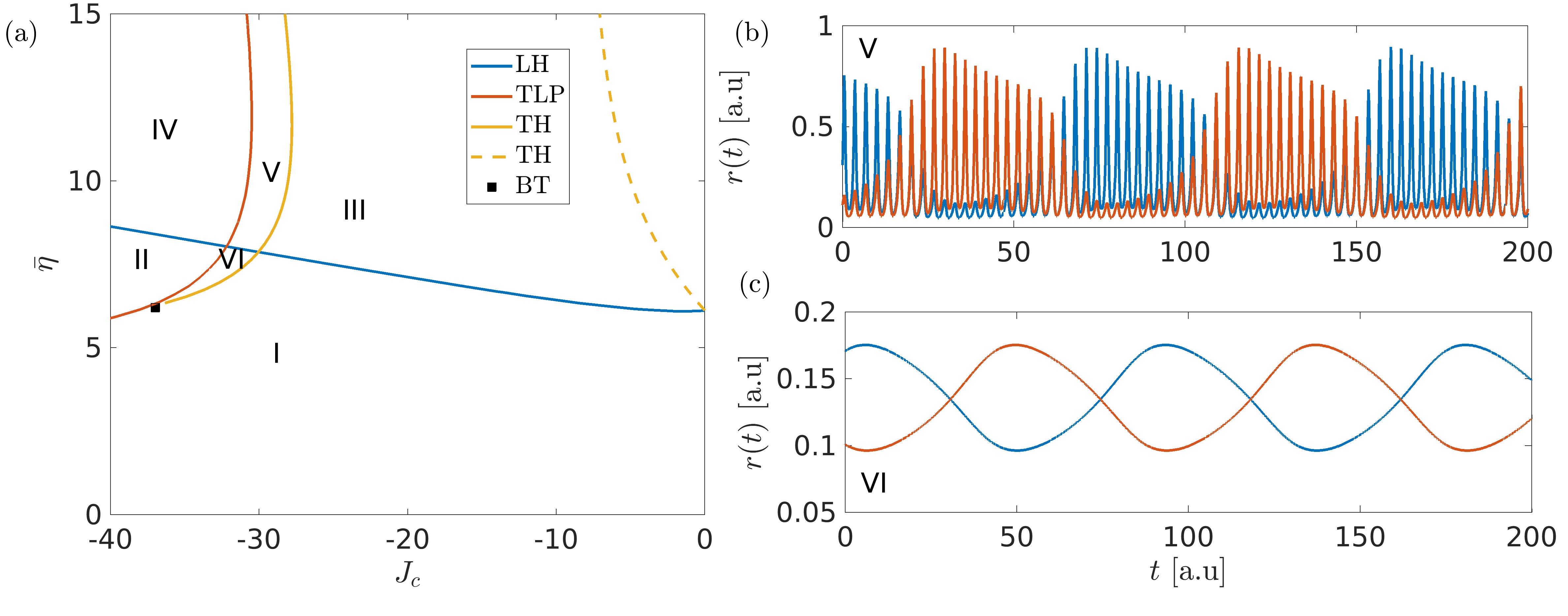}
\caption{{\bf Two symmetrically coupled inhibitory populations with SFA} a) Phase diagram in the $(J_{c},\bar{\eta})$ plane 
depicting the longitudinal (transverse) stability boundary with red (blue) lines. Stability boundaries of stable (unstable)
solutions are shown as solid (dashed) lines. Longitudinal and transverse bifurcation curves divide the 
parameter space in 6 relevant regions. 
b) Typical dynamical evolution in region V: slow-fast nested collective oscillations in anti-phase in the two
populations for $(J_{c},\bar{\eta})=(-32,10)$.
c) Typical dynamical evolution in region VI: antiphase collective oscillations emerging for
$(J_{c},\bar{\eta})=(-32,7.5)$. 
The values of the other parameters are $\tau_s = 2$, $\Delta = 0.5$, $J_{s} = -20$, $\alpha=5$.}
\label{fig:symmetry_inhibitory_adaptation}
\end{figure*}

\subsubsection{Two inhibitory populations}

Let us first analyze the phase diagram in the $(J_c,{\bar \eta})$ plane for two coupled
inhibitory populations with SFA. This is shown in Fig. \ref{fig:symmetry_inhibitory_adaptation} (a)
and it is quite similar to the one obtained in the absence of SFA in Fig. \ref{fig:symmetry_break_alpha0_inh} (a). 
In particular, we observe the same four dynamical regions I-IV previously reported for the case in the absence of adaptation. However, due
to SFA, we have a new transverse Hopf (TH) line (yellow curve in  Fig. \ref{fig:symmetry_inhibitory_adaptation} (a)), which originates from a Bogdanov Takens (BT) bifurcation point.
This new instability leads to the emergence of slow collective oscillations in anti-phase in region VI. An example of this
collective dynamics is reported in Fig. \ref{fig:symmetry_inhibitory_adaptation} (c). By crossing the longitudinal Hopf (LH) instability curve (blue solid line in panel (a)), the slow and fast oscillations, associated to the LH, combine themselves in region V, giving rise to the so-called nested oscillations. In each population these COs are characterized by a slowly varying envelope 
joined to fast oscillations. These oscillations resemble the so-called $\theta$-nested $\gamma$ oscillations observable
in various part of the brain, where the slow modulation is in the $\theta$-range and the fast oscillations in the $\gamma$ band. In particular, these have been measured in the hippocampus of behaving rats and shown to be relevant for cognitive tasks, including navigation, sensory association, and working memory 
\cite{buzsaki1983, colgin2009, buzsaki2012,belluscio2012}. In the present case, due to the symmetry breaking, the two populations
reveal slow-fast nested oscillations in anti-phase as shown in Fig. \ref{fig:symmetry_inhibitory_adaptation} (b).

\begin{figure}
\includegraphics[width=0.95\linewidth]{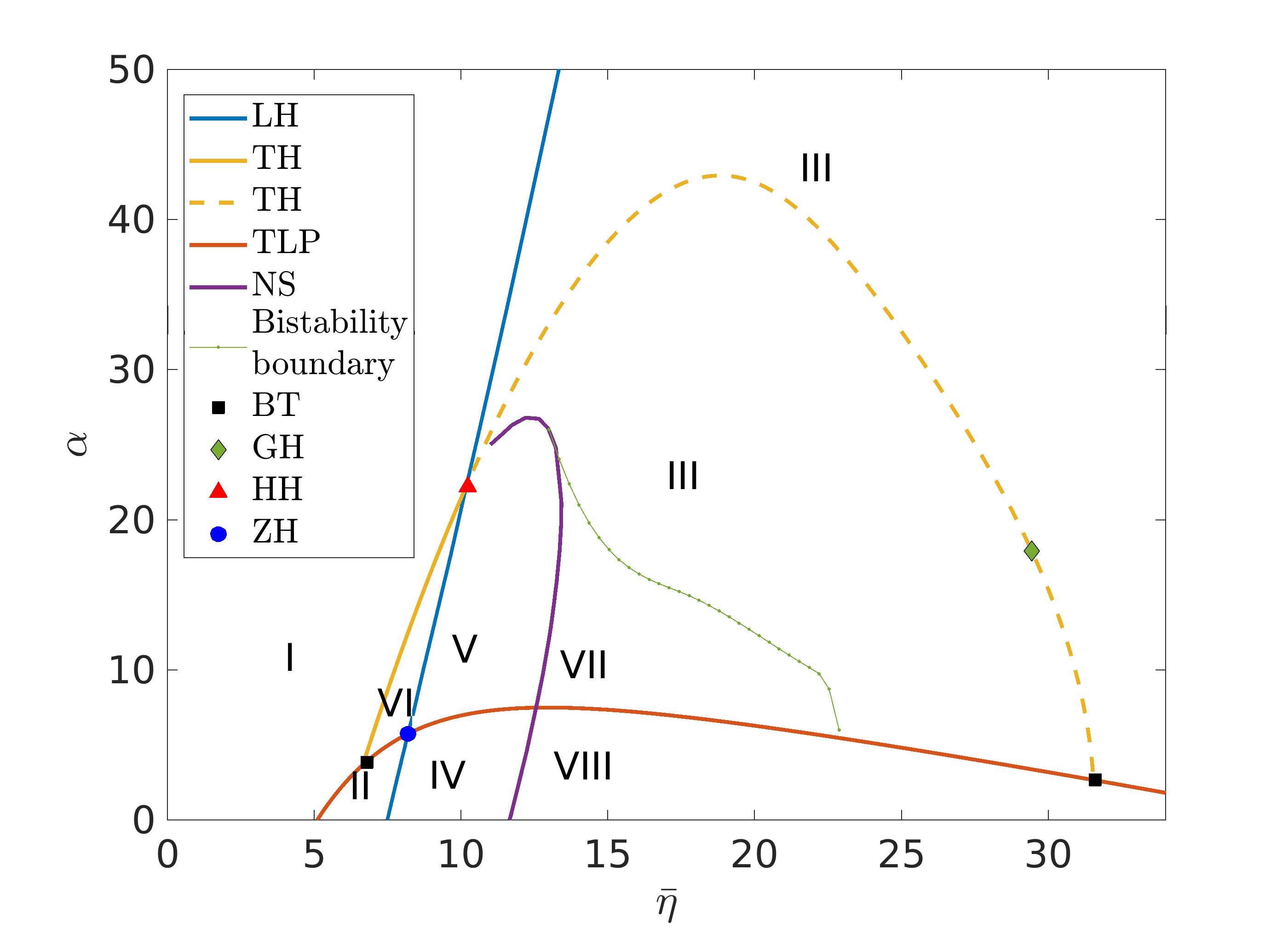}
\caption{{\bf Two symmetrically coupled inhibitory populations with SFA} Phase diagram in the $(\bar{\eta},\alpha)$ plane showing 8 regions with different dynamical regimes
which are listed in details in Table \ref{tab1}.  
Regions I - VI as in Fig. \ref{fig:symmetry_inhibitory_adaptation}; region VII: Coexistence between fast symmetric COs and slow-fast nested COs; region VIII: coexistence between fast
symmetric and asymmetric COs. 
The symbols refer to codimenstion two bifurcation points: namely, Bodgnatov-Takens (BT) (black square),
Generalized-Hopf (GH) (green diamond), Hopf-Hopf (HH) (red triangle), and Zero-Hopf (ZH) (blue circle).
The remaining parameters are fixed as 
$J_{c}=-33$, $J_{s}=-20$, $\Delta = 0.5$, $\tau_A = 10$ and $\tau_s = 2$.
}
\label{fig:symmetry_inhibitory_adaptation_alfa_eta}
\end{figure}

We then proceed to investigate the role played by the adaptation parameter $\alpha$ on the collective dynamics. Therefore, we computed the bifurcation diagram in the $(\bar{\eta},\alpha)$ plane, as shown in Fig. \ref{fig:symmetry_inhibitory_adaptation_alfa_eta}.
Let us first focus on the supercritical longitudinal Hopf (LH) instability (blue solid line), which divides region I and III, where we find
symmetric fixed points and fast symmetric COs, respectively. Here, even for extremely large $\alpha$-values, COs are observable.
This seems in contradiction with the analysis reported for the single inhibitory population with SFA. In such a case, for a constant value of
${\bar \eta}$, the increase of the $\alpha$ parameter led to a shrink of the region where COs were observable, until they disappear completely for sufficiently large $\alpha$ values. However, in the present case, we see that the inhibitory effect, due to the adaptation, can be compensated by increasing the neuron excitability $\bar{\eta}$. Indeed, the positive slope of the LH curve is consistent with this interpretation. 

For low $\alpha$ values we find a transverse instability curve corresponding to a bifurcation line of limit points (TLP) (red solid curve),
which separates symmetric fixed points in region I from asymmetric fixed points in region II. By increasing $\bar{\eta}$ we 
encounter the longitudinal Hopf bifurcation line (LH) previously described, where the asymmetric fixed points bifurcate towards asymmetric fast COs in region IV. 
By following the TLP curve we observe, at a  Bogdanov-Takens (BT) point
located at $(\bar{\eta},\alpha)\approx (6.8,3.8)$, the emergence of a transverse Hopf
(TH) bifurcation line (yellow solid curve in Fig. \ref{fig:symmetry_inhibitory_adaptation_alfa_eta}). This TH line
ends at another Bogdanov-Takens (BT) point observable at very large ${\bar \eta}$ values, always along the TLP curve.
The symmetry breaking associated to the transverse Hopf instability leads from stable symmetric fixed points in region I
to slow COs in antiphase in region VI. By following the TLP curve, we encounter
a Zero-Hopf (ZH) codimension two bifurcation point, when this meets the LH line.
The LH line separates region VI, displaying slow COs in antiphase, 
from region V where slow-fast nested COs in antiphase are observable.

In this phase diagram we can identify two new dynamical regimes
with respect to the ones reported in Fig. \ref{fig:symmetry_inhibitory_adaptation} (a),
both characterized by the coexistence of different stable solutions.
Firstly, let us observe that symmetric fast COs emerge only to the right of
the Neimark-Sacker (NS) curve \cite{NS} (in violet in the figure), that emerges in the proximity of a
Hopf-Hopf (HH) codimension two bifurcation point (red triangle) \cite{HH}.
Indeed, we have found coexisting regimes between symmetric and asymmetric oscillations only to the right of the NS curve.
These regimes exist in region VII, where we have found stable fast symmetric COs together
with stable slow-fast nested COs, and in region VIII, where stable fast symmetric and asymmetric COs coexist. 
The green dotted line in Fig. \ref{fig:symmetry_inhibitory_adaptation_alfa_eta},
separating region VII and III, where only symmetric solutions persist, has been determined by direct simulations of the 
coupled neural masses. 

The last codimension two bifurcation point can be observed along the unstable transverse Hopf branch (yellow dashed line), corresponding to a Generalized Hopf (GH) (green diamond).
It distinguishes between a super-critical Hopf bifurcation line at higher ${\bar \eta}$ values and a sub-critical one at lower ${\bar \eta}$. However, since it involves unstable solutions, these are not observable in the dynamics of the system.



The scenario that we have described  in the $(\bar{\eta},\alpha)$ parameter space is summarized in Table \ref{tab1}.

\begin{table}
 \begin{tabular}{||l|p{6cm}||}
  \cline{1-2}
    Region & Dynamical Regime \\
    \hline
    I &  Symmetric fixed points\\
    \hline
    II &  Asymmetric fixed points\\
    \hline
    III &  Fast symmetric COs\\
    \hline
    IV &  Fast asymmetric COs\\
    \hline
    V &  Slow-fast nested COs in antiphase\\
    \hline
    VI & Slow COs in antiphase\\
    \hline
    VII & Coexistence between fast symmetric COs 
    and slow-fast nested COs in antiphase \\
    \hline
    VIII &  Coexistence between fast symmetric and asymmetric  COs\\
  \cline{1-2}   
\end{tabular}
\caption{Macroscopic dynamical regimes observable in the different regions identified in 
 Fig. \ref{fig:symmetry_inhibitory_adaptation_alfa_eta} }
 \label{tab1}
\end{table}


\begin{figure*}
\includegraphics[width=0.95\linewidth]{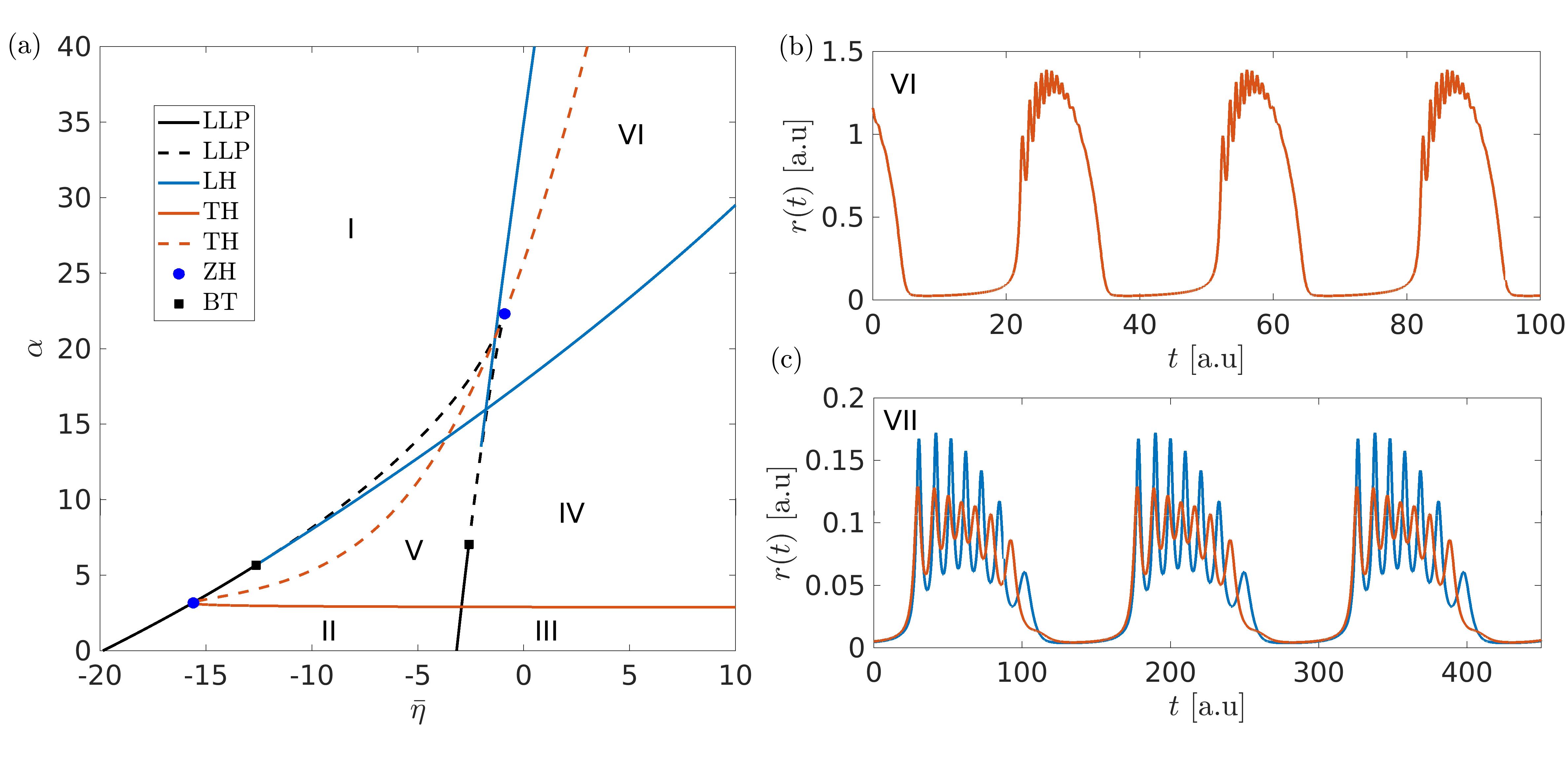}
\caption{{\bf Two symmetrically coupled excitatory populations with SFA} a) Phase diagram in the $(\bar{\eta},\alpha)$ space showing 6 regions with different dynamics. Parameters for this diagram: $J_{c}=20$, $J_{s}=8$, $\Delta = 0.5$, $\tau_A = 10$ and $\tau_s = 2$. Regions I - V as in Fig. \ref{fig:symmetry_break_alpha0_exc} (b). b) Region VI: symmetric bursting dynamics $({\bar \eta},\alpha)=(5,30)$. c) 
An example of bursting with broken symmetry for parameters $({\bar \eta},\alpha)=(23,70)$.
}
\label{fig:symmetry_excitatory_adaptation_eta_alpha}
\end{figure*}

\subsubsection{Two excitatory populations}

Let us now consider two symmetrically coupled excitatory populations with SFA. For not too large
$\alpha$-values we observe the same scenario as the one reported in Fig. \ref{fig:symmetry_break_alpha0_exc} (b)
in the absence of adaptation. Therefore we do not describe in detail the possible solutions here,
since they have been already reported in the previous sub-section \ref{2exc}. 
In order to observe new dynamical regimes due to adaptation, we rather consider
the $(\bar{\eta},\alpha)$ plane for sufficiently large $\alpha$-values. Indeed, as shown in
Fig. \ref{fig:symmetry_excitatory_adaptation_eta_alpha} (a), a new region VI emerges where identical bursting oscillations can be observed in both populations, as shown in panel (b). 
This region is delimited by two LH lines separating it from region I, where the fixed point, corresponding to low firing activity, is stable, 
and region IV, where fast COs in anti-phase are observable. The low activity state gets unstable by crossing 
the almost vertical LH line; once crossed the LH line, it gives rise to symmetric relaxation oscillations corresponding to a burst  characterized by a silent regime connected to fast modulated oscillations. This bursting dynamics is analogous to the one  reported for a single excitatory population with SFA, shown in Fig. \ref{fig:Hopf_boundary_J_eta} (b).
The other LH line determines the stability of the COs in anti-phase, which emerge
via a transverse Hopf (TH) bifurcation (red solid line) from the destabilization of the high activity
fixed point present in region III.
 
As observable in Fig. \ref{fig:symmetry_excitatory_adaptation_eta_alpha} (a), for low $\alpha$ values and high negative $\bar{\eta}$ values, the system shows only two possible stable regimes separated by a longitudinal saddle node bifurcation curve (black solid line): a low activity fixed point in region I and a coexistence between 
two fixed points with different activity levels in region II. Further increasing $\alpha, \bar{\eta}$ along the branch point curve (LLP) leads to a codimension two ZH point. 
Here emerges a new longitudinal Hopf curve (blue solid line) which delimites the region V, where
a coexistence between anti-phase COs and a low activity fixed point is present.
As we have already shown, anti-phase COs are present also in region IV, as the unique
stable solution. Beyond these broken symmetry solutions, where the same attractor is visited
at different times by the two populations, we have been able to identify another solution with broken symmetry at quite large $\alpha$-values,
as shown in  Fig. \ref{fig:symmetry_excitatory_adaptation_eta_alpha} (c).
In this case, both populations display bursting activity, characterized by the same period of the slow modulation, but with 
different amplitudes and frequencies for the modulated fast spiking oscillations.

\section{$\theta$-$\gamma$ cross-frequency coupling due to adaptation}


In this Section we will focus on the dynamical regimes where 
cross-frequency couplings among slow and fast collective oscillations
are observable for two coupled excitatory or inhibitory populations with SFA.
In particular,  due to their relevance for brain dynamics  we
will consider as slow rhythms the ones in the $\theta$ range 
(1-12 hz) and as fast ones COs in the $\gamma$ band
(30 - 130 Hz). 

\subsection{$\theta$-nested $\gamma$ oscillations in coupled inihibitory populations with SFA}

\begin{figure*}
\includegraphics[width=0.95\linewidth]{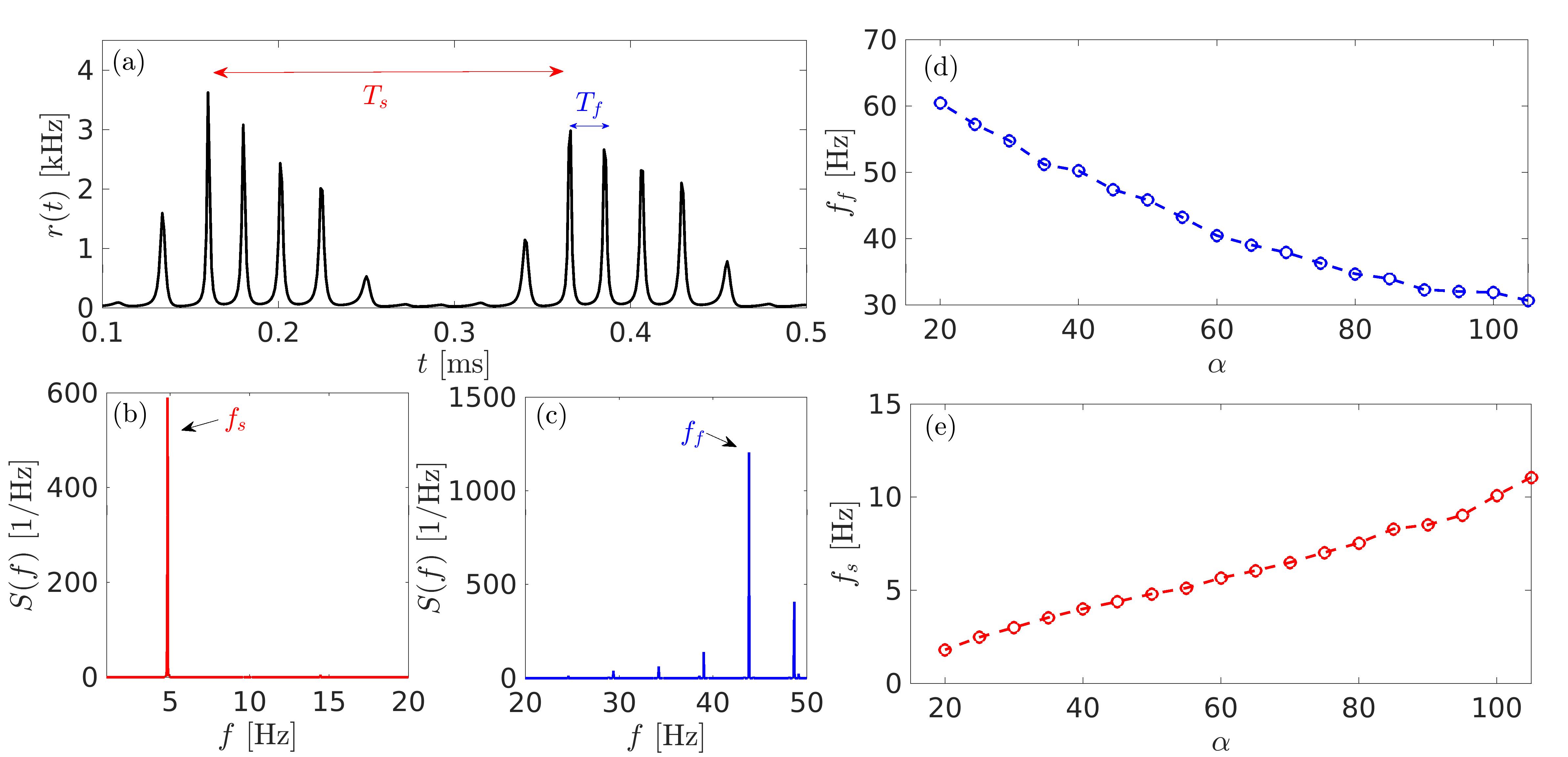}
\caption{{\bf Two coupled inhibitory populations with SFA}
a) Time trace of the firing rate $r(t)$ for one of the two populations. The other population shows the same evolution, but shifted in phase.
The two relevant periods of the oscillatory behavior are reported as 
$T_s$ (slow) and $T_f$ (fast). b) Power spectral density $S(f)$ of the time trace filtered using a 3rd order Butterworth low pass filter with cut-off frequency of
15 Hz. c) Same as in b) using a 3rd order Butterworth high pass filter with cut-off frequency of 20Hz. d)-e) Variation of the fast and slow frequency components $f_f$ and $f_s$, respectively, as a function of the adaptation parameter $\alpha$. Parameters used here: $\bar{\eta} = 28$, $J_{s}=-20$, $J_{c}=-47$, $\Delta=0.5$, $\tau_S = 20$ms and $\tau_A = 100$ms. For panels (a-c), a value of $\alpha = 50$ was chosen.}
\label{fig:nested_frequency_alpha}
\end{figure*}

For two coupled inhibitory populations with SFA the slow-fast nested COs in anti-phase shown in Fig. \ref{fig:symmetry_inhibitory_adaptation} (b)
can be classified as instances of CFCs. Here we will consider parameters allowing the slow (fast) rhythms to be in the $\theta$ ($\gamma$) range
and we will analyze the dependence of these rhythms on the adaptation. In this case
the phase of the $\theta$ modulation affects the amplitude of the $\gamma$-oscillations: 
this CFC mechanism is termed phase amplitude coupling and it turns out to have been observed in several parts of the brain \cite{canolty2010}. Since the time traces of the two populations
are identical, but in anti-phase, we limit the analysis to only one trace. As shown in Fig. \ref{fig:nested_frequency_alpha} (a),
one observes distinct oscillatory periods: a slow inter-cycle period $T_s$, which identifies the time in between two successive $\theta$ oscillations, and a fast intra-cycle period $T_f$ 
between successive gamma peaks.

The evaluation of the power spectral density $S(f)$ for the time trace reported in
Fig. \ref{fig:nested_frequency_alpha} (a) reveals a single peak at low frequencies $f_s \simeq 5$ Hz
(see panel (b)) and a main peak at high frequencies, located around $f_f \simeq 35$ Hz, plus smaller
peaks in correspondence of the combinations of the 2 main frequencies $f_f \pm k f_s$ with $k=1,2,3,\dots$ (as shown in panel (c)). 

It is of particular interest to understand how the parameter $\alpha$, measuring the strength 
of the adaptation, influences these two frequencies. The analysis is reported in Fig. \ref{fig:nested_frequency_alpha} (d) and (e) for the chosen set of parameters and for $\alpha \in [20,100]$.
It is evident that $f_f$, that identifies the main peak in the power spectrum of the fast signal component, decreases monotonously for increasing $\alpha$, while the main peak of the slow signal component $f_s$ increases. In particular, $f_s$ grows from 2 to 11 Hz, while the fast peak $f_f$ decreases from 60 to 30 Hz, thus exploring the whole range of the so-called slow $\gamma$ rhythms observable
in the  hippocampus \cite{colgin2009}.

\subsection{$\theta$-$\gamma$ population bursts in coupled excitatory populations with SFA}

\begin{figure*}
\includegraphics[width=0.95\linewidth]{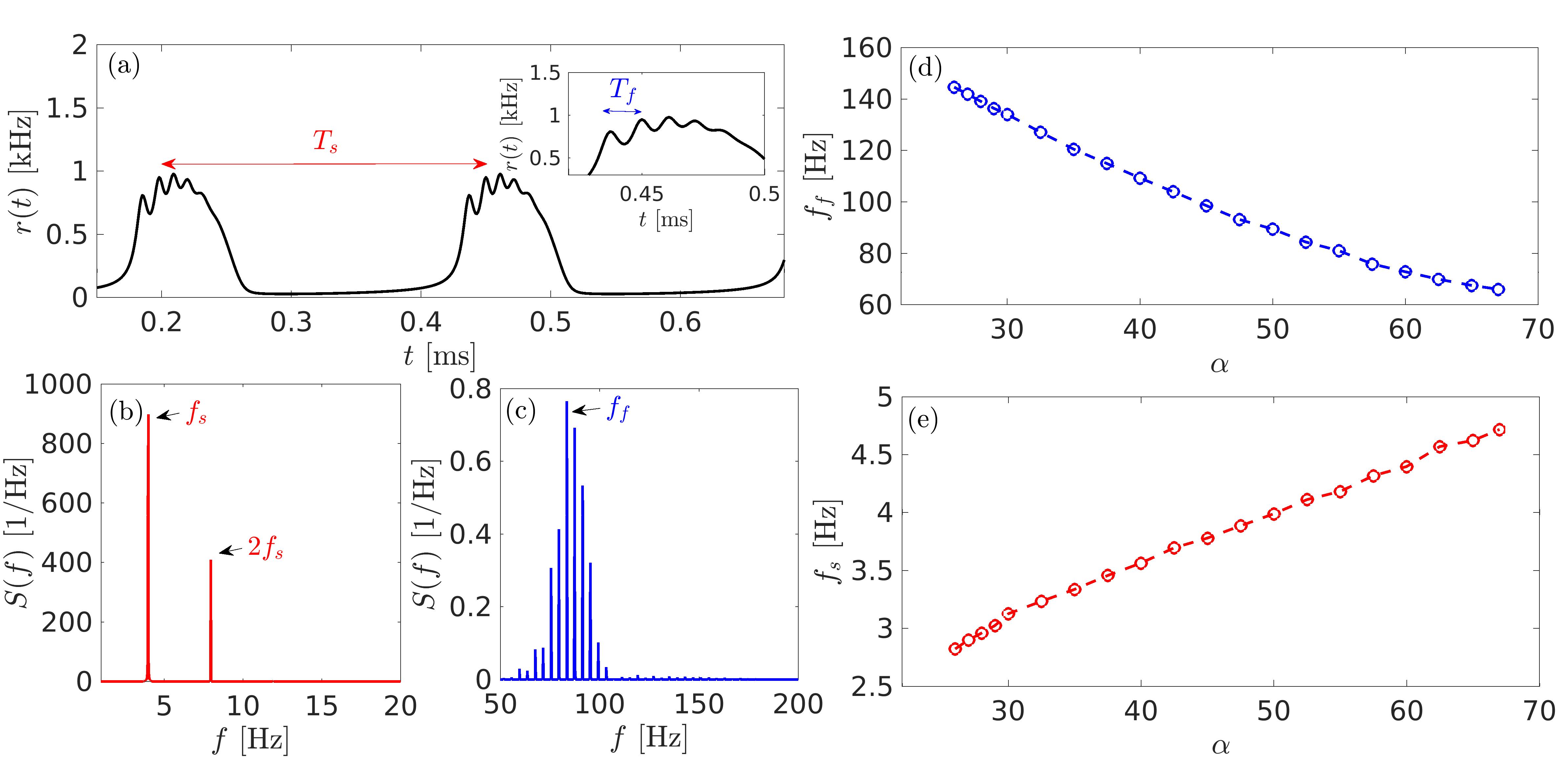}
\caption{{\bf Two coupled excitatory populations with SFA}
a) Time trace of the firing rate $r(t)$ for one of the two populations. The other population shows exactly the same evolution.
The two relevant periods of the oscillatory behavior are sketched as $T_s$ (slow) and $T_f$ (fast) (for this see the enlargement in the inset).
b) Power spectral density $S(f)$ of a filtered time trace obtained by using a 3rd order Butterworth with cut-off frequency of
10Hz. c) Same as in (b) by using a 3rd order Butterworth high pass filter with cut-off frequency of 80 Hz. d)-e) Variation of the fast and slow frequency components $f_f$ and $f_s$, respectively, as a function of the adaptation constant $\alpha$. Parameters used here: $\bar{\eta} = 5$, $J_{in}=8$, $J_{ex}=20$, $\Delta=0.5$, $\tau_S = 20$ms and $\tau_A = 100$ms. For panels (a-c), a value of $\alpha = 50$ was chosen.}
\label{fig:bursting_frequency_alpha}
\end{figure*}

For coupled excitatory neural masses with SFA it is possible to observe
CFC in the form of a burst connecting a resting state to a fast spiking activity.
In this case the dynamics is the same for both populations.
For specific choices of the parameters the inter-burst period is in the $\theta$-range,
while the intra-bursts are $\gamma$ oscillations (an example is reported in  Fig. \ref{fig:bursting_frequency_alpha} (a)).
Analogously to the previous sub-section, we can define $T_s$ as the inter-burst period, while $T_f$ represents the intra-burst period, as shown in Fig. \ref{fig:bursting_frequency_alpha} (a). 
By performing  low-pass (high-pass) filtering of the time signal reported in the above mentioned figure (i.e. the firing rate $r(t)$ of one of the two populations)  we can detect 
the main peak $f_s \simeq 4$ Hz ($f_f \simeq 80$ Hz) for the slow (fast) component of the power spectrum $S(f)$,
see Fig. \ref{fig:bursting_frequency_alpha} (c) and (d).
At variance with the previous case, the phase of the $\theta$ rhythm does not modulate
only the amplitude of the $\gamma$-oscillations, but also their periods (frequencies), therefore the
spectrum reported in panel (c) reveals that an entire band of frequencies between 70 and 100 Hz is indeed
excited.

Also in the present case, we analyze the dependence of $f_s$ and $f_f$ on the adaptation parameter $\alpha$.
By varying $\alpha$ in the range $[25,68]$, we observe that $f_s$ ($f_f$) increases (decreases)
for growing values of $\alpha$ as before. However, while the variation of $f_s$ is quite limited from 2.8 to 4.8 Hz, the gamma peak $f_f$
ranges from 130 to 60 Hz, i.e. $f_f$ is now exploring the fast $\gamma$ range \cite{colgin2009}.

\section{Discussion}

In this paper we have analyzed the influence of the spike frequency adaptation mechanism on the macroscopic
dynamics of single as well as symmetrically coupled neural populations connected via  either excitatory or inhibitory
exponentially decaying post-synaptic potentials. The macroscopic dynamics has been investigated
in terms of next generation neural masses reproducing exactly the dynamics of fully coupled networks of
quadratic integrate-and fire spiking neurons in the thermodynamic limit. This low dimensional mean field reduction has allowed for the estimation of the corresponding bifurcation diagrams. In particular, we 
mostly focused on the instabilities of the stationary states with emphasis on the longitudinal and transverse nature of the instabilities in symmetrically coupled populations.

In a single population with a fixed distribution of the neural excitabilities, SFA favors the emergence of population bursts when the coupling is excitatory, while it hinders the emergence 
of periodic population spikes in inhibitory networks. 
This is related to the fact that SFA acts as an effective inhibition with a slow timescale $\tau_a$.
In inhibitory networks in the absence of adaptation, oscillations are already present, controlled by the fast synaptic time scale $\tau_s$.
Therefore, in this case, SFA leads to an additional inhibitory effect that
can eventually kill the population spiking for sufficiently large amplitude of the adaptation parameter
\cite{devalle2017firing, ceni2020}. On the other hand, in excitatory networks
COs are absent without adaptation, while these are promoted by the 
SFA, which represents the only source of inhibition.
As we have shown by performing a slow-fast decomposition, the population burst evolves on a time scale controlled by $\tau_a$ connecting
a low firing state to a high firing one, that is approached via fast damped population spikes characterized by a fast time scale of the order of $\tau_s$.

The coupling of two identical populations in a symmetric way leads to the 
emergence of collective solutions with broken symmetry for cross-couplings larger than self-couplings
($|J_c| > |J_s|$). These solutions can be classified as chimera-like for the inhibitory case
and anti-phase population spikes in the excitatory one. The chimera-like solutions correspond
to asymmetric fixed points and fast COs. These kind of solutions have been usually identified whenever the self-coupling was larger than the cross-coupling in identically coupled populations of 
phase oscillators \cite{abrams2004} and excitatory neurons \cite{olmi2011, ratas2017symmetry}.
In this case the action of the phase oscillators (neurons) in their own population was more effective in promoting 
the synchronized (firing) activity than that of the oscillators (neurons) in the other population.
This scenario would be observable for two inhibitory coupled populations whenever $|J_c| > |J_s|$. Therefore, 
consistently with our analysis, we can affirm to have reported, for the first time to our knowledge,
chimera-like states in purely inhibitory neural systems.

Chimera-like states have been reported also in  \cite{ratas2017symmetry} for two coupled next generation
neural masses interacting via steps of currents for the excitatory and inhibitory case. However it should be noticed that
in \cite{ratas2017symmetry}, the inhibitory case refers to a situation where the self-coupling is excitatory,
while the cross-coupling is inhibitory. This choice violates the Dale's principle \cite{dale1934}
that states that neurons can transmit either excitatory or inhibitory post-synaptic signals, but not 
both at the same time. However a set-up like the one proposed in \cite{ratas2017symmetry}, could
be shown to mathematically represent two excitatory populations coupled to an inhibitory pool \cite{wong2006},
thus resulting in a set-up involving at least three neural populations and definitely different from the one we have examined.

The addition of SFA leads to new macroscopic solutions and regimes besides those observed in the absence of adaptation. 
In the inhibitory case, new collective solutions with broken
symmetry are observable, that correspond to slow collective oscillations and 
slow-fast nested oscillations, both emerging in anti-phase between the two populations. 
Furthermore, the fast symmetric COs can coexist with fast asymmetric ones
as well as with slow-fast nested COs in antiphase. The new solutions
observable for excitatory populations with SFA correspond to population bursts
that are either symmetric or asymmetric.
 
The evolution of two symmetrically coupled neural masses with SFA
can lead to interesting CFC phenomena, where oscillations 
induced both by the fast synaptic time scale and the slow adaptation
are present at the same time and interact among them. 
In particular, we have examined macroscopic solutions exhibiting
cross-frequency coupling in the $\theta$-$\gamma$ range and the dependence
of the fast $f_f$ and slow $f_s$ frequencies on the adaptation, both for the
inhibitory and excitatory case.

As a first result, we observe that the reduction of SFA leads to an increase
in the frequency  of the $\theta$-rhythm, joined to a decrease in the $\gamma$
frequencies. This is analogous to the effect of cholinergic modulation, 
as shown in several experiments based on EEG recordings of the hippocampus and the olfactory cortex 
\cite{konopacki1987,biedenbach1966}. In particular, cholinergic modulation induces an increase in the number of $\gamma$ oscillations, joined
to a decrease in the frequency of the $\theta$-rhythm \cite{traub1992, barkai1994, crook1998}.
Indeed, consistently with these results, it has been shown that cholinergic drugs are responsible for a  reduction of SFA 
in pyramidal cells  \cite{aiken1995}.

Furthermore, it has been shown that $\theta$-$\gamma$ CFC is favoured by the presence of the
cholinergic neurotransmitters \cite{newman2013,howe2017} and, therefore, by a reduction of
SFA \cite{yang2021}. This is somehow consistent with our results, since we have shown that the ratio
$f_f/f_s$ decreases for increasing $\alpha$, thus determining a reduction of the number 
of $\gamma$ oscillation events inside a single $\theta$ cycle for higher adaptation values.
Therefore, according to the hypothesis that the neural code for multi-item messages is organized by 
$\theta$-$\gamma$ nested brain oscillations \cite{lisman2013, lisman2008}, 
we expect that, for large adaptation values, the number of coded items will diminish,
thus reducing the neural coding  effectiveness.
 
Finally, we have shown that, in our model, $\theta$ rhythm can modulate either
slow or fast $\gamma$ oscillations, analogously to those identified in the hippocampus and other 
brain areas \cite{colgin2016}. In particular, we observe that, in presence of SFA, slow $\gamma$ are associated to 
inhibitory dynamics, while excitatory populations support fast $\gamma$ oscillations.

Our analysis could be useful to develop new models of Central Pattern Generators, 
which are responsible for the generation of rhythmic movements, since these
models are often based on two interacting oscillatory populations with adaptation,
as reported for the spinal cord \cite{linden2021} and the respiratory system \cite{rubin2011}.
Finally, the model of the medullary circuitry reported in \cite{deschenes2016} can be taken as a cue for a 
further interesting application of coupled neural masses with SFA, like the modelization of rhythmic 
whisking in rodents and, in particular, the entrainment of whisking by breathing.

\section*{Appendix A: Stability of the stationary solutions for a single population}

Here we show the steps to calculate the stability boundaries
for the stationary solutions of the single population. We start by using the adimensional system defined in Eq.
 \eqref{eq:mean_field2}. The characteristic polynomial evaluated at
the equilibrium $({r}_0,{v}_0,{s}_0,{a}_0)$ is given by
\begin{equation}
\label{eq:charact_polynomial}
p(\lambda)=\frac{2\alpha {r}_0 \Lambda_s+\Lambda_a (-2J {r}_0+(4\pi^2 {r}_0^2 + (\lambda-2 {v}_0)^2) \Lambda_s)}{\tau_a \tau_s}
\end{equation}
with $\Lambda_a = (1+\lambda \tau_a)$ and $\Lambda_s = (1+\lambda \tau_s)$.
In general a bifurcation occurs when one or more eigenvalues cross the imaginary axis, namely whenever $\lambda = i \omega$, hence we look for solutions of the form $p(i \omega) = 0$. Replacing $\lambda$ with $i \omega $ in \eqref{eq:charact_polynomial}
and equating to zero leads to 
\begin{equation}
\label{eq:Poly_Omega_i}
2\alpha {r}_0 \Omega_s+\Omega_a (-2J {r}_0+(4\pi^2 {r}_0^2 + (2 {v}_0-i\omega)^2) \Omega_s)=0,
\end{equation}
with $\Omega_s = (1 + i \omega \tau_s)$ and $\Omega_a = (1 + i \omega \tau_a)$. For Eq. \eqref{eq:Poly_Omega_i} to be satisfied, both 
imaginary and real parts of $p(i\omega)$ should be zero, namely:
\begin{equation}
\label{eq:Re_Im_Poly}
\text{Re}[p(i \omega)] = 0 \enskip;\enskip \text{Im}[p(i \omega)] = 0 \enskip.
\end{equation}

By solving $\text{Re}[p(i \omega)] = 0$ ($\text{Im}[p(i \omega)] = 0$)
in \eqref{eq:Re_Im_Poly} we get the solutions
$\omega^*_{Re}$ ($\omega^*_{Im}$). These have the form:
\bes
\begin{eqnarray}
\label{eq:omegaRe} \omega_{Re}^* &=& \pm \frac{\sqrt{P_1 \pm \sqrt{P_2+P_1^2}}}{\sqrt{2\tau_a \tau_s}} \\
\label{eq:omegaIm} \omega_{Im}^* &=& \pm \sqrt{2P_3}, \quad \omega^*_{Im}¨ = 0 
\end{eqnarray}
\ees

with

\begin{widetext}
\bes
\begin{eqnarray}
P_1 &=& 1+4\pi^2 {r}_0^2\tau_a \tau_s+4 {v}_0^2\tau_a\tau_s - 4 {v}_0(\tau_a+\tau_s) \\
P_2 &=& 8(J {r}_0-2 {r}_0^2- {r}_0(2\pi^2 {r}_0+\bar{\alpha}))\tau_a\tau_s \\
P_3 &=& \frac{2 {r}_0-2 {r}_0^2(\tau_a + \tau_s)- {r}_0(-J\tau_a + \alpha \tau_s + 
2\pi^2 {r}_0(\tau_a + \tau_s))}{-\tau_s+\tau_a(-1+4 {v}_0\tau_s)}
\end{eqnarray}
\ees
\end{widetext}

Recalling that ${v}_0 = -\Delta / (2\pi {r}_0)$, all the expressions are parametrized by the coordinate ${r}_0$ of the equilibrium.
Equating \eqref{eq:omegaRe} and the non zero solutions of \eqref{eq:omegaIm}, and solving for one of the parameters,
it is possible to parametrize the Hopf bifurcation solution via ${r}_0$.
Similarly, equating \eqref{eq:omegaRe} with the trivial solution  $\omega^*_{Im}=0$, the Limit point type bifurcations (saddle-node, pitchfork and transcritical) can be obtained.

\acknowledgments
We acknowledge useful interactions with Gloria Cecchini.
AT received financial support by the Labex MME-DII (Grant No ANR-11-LBX-0023-01) and by the ANR Project ERMUNDY (Grant No ANR-18-CE37-0014), all part of the French programme ``Investissements d'Avenir''. Part of this work has been developed during the visit of SO during 2021 to the Maison internationale de La Recherche, Neuville-sur-Oise, France supported by CY Advanced Studies, CY Cergy Paris Universit\'e, France.

\bibliographystyle{apsrev4-1}

\bibliography{biblioPaper}

\end{document}